\documentstyle[12pt]{article}

\newcommand{\EQ}{\begin{equation}}
\newcommand{\EN}{\end{equation}}
\newcommand{\bea}{\begin{eqnarray}}
\newcommand{\ena}{\end{eqnarray}}

\begin{document}
 \def\bq{\begin{quote}}
\def\eq{\end{quote}}
\topmargin -1.2cm
\oddsidemargin 5mm

\renewcommand{\Im}{{\rm Im}\,}
\renewcommand{\thefootnote}{\fnsymbol{footnote}}
\newcommand{\sect}[1]{\setcounter{equation}{0}\section{#1}}
\renewcommand{\theequation}{\thesection.\arabic{equation}}

\newpage
\begin{titlepage}
\begin{flushright}
IFUM 484/FT\\
July 1995 \\
\end{flushright}
\vspace{1.5cm}\begin{center}
{\bf{{\large RENORMALIZATION OF MATTER FIELD THEORIES } \\
{\large ON THE LATTICE AND THE FLOW EQUATION}}}
\footnote{Work supported in part by M.U.R.S.T. and
EEC, Science Project SC1${}^*$-CT92/0789.}\\
\vspace{1cm}
{ M. PERNICI}  \\
\vspace{2mm}
{\em INFN, Sezione di Milano, Via Celoria 16, I-20133 Milano, Italy}\\
\vspace{4mm}
{ M. RACITI and F. RIVA}\\
\vspace{2mm}
{\em Dipartimento di Fisica dell'Universita' di Milano, I-20133 Milano,
Italy}\\
\vspace{2mm}
{\em and INFN, Sezione di Milano, Via Celoria 16, I-20133 Milano, Italy}\\
\vspace{0.8cm}
\vspace{2cm}
{{\bf{ABSTRACT}}}
\end{center}
\bq

We give a new proof of the renormalizability of a class of
matter field theories on a space-time lattice; in particular we consider
$\phi^4$ and massive Yukawa theories with Wilson fermions.
We use the Polchinski approach to renormalization, which is based on
the Wilson flow equation; this approach is substantially simpler than the
BPHZ method, applied to the lattice by Reisz.

We discuss matter theories with staggered
fermions. In particular we analyse a simple kind of staggered fermions
with minimal doubling, using which
we prove the renormalizability of a chiral sigma model
with exact chiral symmetry on the lattice.

\eq
\vfill
\end{titlepage}
\renewcommand{\thefootnote}{\arabic{footnote}}
\setcounter{footnote}{0}

\newpage

Perturbative quantum field theories on the lattice \cite{w1} present a
few difficulties that are absent in other regularization schemes;
their renormalization has been
studied by Reisz \cite{r1,r2,r3} in the BPHZ approach \cite{b1,r4}.
Beyond the usual technical difficulties of the BPHZ approach, requiring
an analysis of the topological structure of the Feynman graphs, there
are those specific to lattice theories, due to the periodicity
of the Green functions in
momentum space. In the first place, it is non-trivial to prove a
power-counting theorem
on the lattice \cite{r1}, since the Feynman integrands
are not rational functions of the momenta, but of trigonometric functions
of the momenta.
In the second place one has to show that
the subtraction procedure can be implemented with
counterterms which are local on the lattice.

In lattice gauge theories there is an additional difficulty, the
presence of an infinite number of irrelevant interaction terms in
the bare action; Reisz \cite{r3} has shown that these vertices do not
modify the renormalized Green functions computed in the continuum.

In theories containing fermions
there is the doubling problem for fermions on the lattice \cite{n1,k1};
introducing the Wilson term \cite{w2} in the fermionic action
the doublers decouple in the continuum
limit, but chiral symmetry is broken in a hard way by the Wilson term.
Reisz's results apply to Wilson fermions \cite{r1,r2,r3,l1}; these results
have not been extended to fermionic models with doublers. While
chiral invariance is not maintained by Wilson fermions, it is possible
to construct models with doublers having chiral invariance on the
lattice, for instance with the Kogut-Susskind \cite{s1}
staggered fermions,
or with simpler versions of staggered fermions \cite{w4,p2};
it has not yet been shown that these models are renormalizable.

A simple approach to renormalization in continuum quantum field
theory has been initiated by Polchinski \cite{p1}, in the spirit of
the Wilson renormalization group \cite{w3}; this approach has been further
simplified and improved in \cite{k2,BM,b2}.
One studies
an effective action which, besides the ultraviolet cut-off $\Lambda_0$,
has an `infrared' cut-off $\Lambda$. The effective action satisfies
perturbatively a
linear differential equation in $\Lambda$, called the flow equation.
One can impose mixed boundary
conditions on the effective action, fixing the relevant terms of the
effective action at a renormalization scale $\Lambda = \Lambda_1$,
while the irrelevant terms are fixed at a scale
$\Lambda = \Lambda_0$.

In this paper we study the renormalizability of scalar and Yukawa theories
on the lattice; we adapt the Polchinski approach to the lattice, introducing
an `infrared' cut-off $\Lambda$, which restricts the propagators on a
periodic band in momentum space,
at a distance of the order of $\Lambda$ from the poles of the propagators.

We extend Reisz's proof of renormalizability to a class of Yukawa theories
with a simple kind of staggered fermions \cite{w4,p2}.

In the first section we study the massive $\phi^4$  field theory
on a hypercubic lattice;
we prove that it is multiplicatively renormalizable, using the flow
equation for the amputated connected  Green functions; we show that
the addition of irrelevant terms to the action does not change
the renormalized Green functions.

In the second section we prove the renormalizability of a general
class of matter field theories on a lattice including scalars,
fermions and auxiliary fields.

In the third section we apply these results to
Wilson fermions and a simple kind of staggered
fermions with minimal doubling; using
the latter we prove the renormalizability of
a chiral sigma model with explicit chiral invariance on a hypercubic lattice.
The proof is obtained by rewriting the model on a reduced lattice, on which
the results of the previous section apply.

\vskip 2.0 cm

\sect{Renormalization of the massive $\phi^4$ model on a hypercubic lattice.}

\vskip 1.0 cm

{\bf 1.1 Introduction.}

Consider an infinite space-time hypercubic lattice with lattice spacing
$a$ and sites
in $x_{\mu} = n_{\mu} a$, where $n_{\mu}$ are integers, $\mu = 1,...,4$.
The bare action is
\bea
S = \frac{a^2}{2} \sum_{x,{\mu}} \phi^{(0)}_x
( 2 \phi^{(0)}_x - \phi^{(0)}_{x + \hat{\mu}}
- \phi^{(0)}_{x - \hat{\mu}} ) +
a^4 \sum_x ( \frac{m^{(0)2}}{2}\phi^{(0)2}_x +
\frac{g^{(0)}}{4!} \phi^{(0)4}_x ) + S_{irr}
\ena
where $\phi^{(0)}_x$ is the bare field; $m^{(0)}$ and $g^{(0)}$ are the
bare parameters.  $S_{irr}$ is an irrelevant term, which will be
specified below; it will be chosen in such a way that
the action preserves the hypercubic
space-time symmetry and the discrete symmetry
$\phi^{(0)}_x \rightarrow - \phi^{(0)}_x$.

 The action reads, in momentum space,
\bea
S = \frac{1}{2} \int_k \phi^{(0)} (-k) ( \hat{k}^2 + m^{(0)2} ) \phi^{(0)} (k)
+ \frac{g^{(0)}}{4!} \int_{k_1,...,k_4} (2 \pi)^4
\delta_P^4( \sum_1^4 k_i) \prod_{i=1}^4\phi^{(0)} (k_i)
+ S_{irr}
\ena
where $\hat{k}_{\mu} = \frac{2}{a} sin \frac{k_{\mu} a}{2}$,~~
$\delta_P^4(k) = \frac{a^4}{(2\pi)^4} \sum_x e^{-ikx}$~ is the
periodic delta function,

\noindent
$\int_k \equiv \int_{-\frac{\pi}{a}}^{\frac{\pi}{a}} \frac{d^4k}{(2\pi)^4}$.

The Fourier transform of $\phi_x$ is
$\phi (k) = \frac{a^4}{(2\pi)^4} \sum_x e^{-ikx} \phi_x$, and
$\phi_x = \int_k \phi(k) e^{i kx}$.

The Fourier transform $\phi (k)$ and the kinetic operator $\hat{k}^2$
are periodic under $k_{\mu} \rightarrow k_{\mu}+\frac{2\pi}{a}$.
The momenta can be restricted to the Brillouin zone
$|k_{\mu}| \leq \frac{\pi}{a}$, provided one requires momentum
conservation at the vertices only modulo $\frac{2\pi}{a}$.

\bea
\Lambda_0 \equiv \frac{\pi}{a}
\ena
is a kind of ultraviolet cut-off on the cartesian components of the
momenta.

In terms of the renormalized field
$\phi_x = Z^{- \frac{1}{2}} \phi^{(0)}_x$
the action becomes
\bea
S[\phi] = \frac{1}{2} \int_k \phi (-k) ( \hat{k}^2 + m^2 ) \phi (k)
+ S_I[\phi]
\ena
with
\bea
S_I[\phi] =
\frac{1}{2} \int_k \phi (-k) (c_1 + c_2 \hat{k}^2 ) \phi (k)
+ \frac{c_3}{4!} \int_{k_1,k_2,k_3,k_4} (2\pi)^4
\delta_P^4( \sum_1^4 k_i) \prod_{i=1}^4 \phi (k_i) + S_{irr}[\phi]
\ena
where
\bea
c_1 = Z m^{(0)2} - m^2 ~~;~~c_2 = Z - 1 ~~;~~ c_3 = g^{(0)} Z^2
\ena
In terms of the loop expansion parameter $\hbar$ the renormalization constants
$c_i$ have the series expansion
$c_i = g \delta_{i,3} + \sum_{l=1}^{\infty} \hbar^l c^{(l)}_i$.
The $c^{(l)}_i$'s are constants depending on $\Lambda_0$, $m, g$ and on
the renormalization conditions. A standard set of renormalization
conditions is, in terms of the proper vertices,
\bea
\Gamma_2|_{p=0} = m^2 ~~~~;~~~~\frac{1}{8}
\sum_{\mu = 1}^4 \frac{\partial^2 \Gamma_2}
{\partial p_{\mu} {\partial p_{\mu}}}|_{p=0} = 1 ~~~~; ~~~~
\Gamma_4|_{p=0} = g
\ena
The parameters $m$ and $g$ are called renormalized
parameters. A more general set of renormalization conditions will
be specified later.

$S_{irr}[\phi]$ contains irrelevant terms which are defined on the
lattice; it will be chosen linear
in a parameter $y \in [0,1]$, $S_{irr}[\phi] = y S_{irr}[\phi]|_{y=1}$.
The dependence of $S_{irr}[\phi]$ and of the Green functions on $y$
will not be explicitly indicated in the following.
A typical choice (though not the most general) is
$S_{irr}[\phi] = y \sum_x \sum_{n=3}^{\infty} d_{2n} a^{2n} \phi_x^{2n}$,
where $d_{2n}$, with $n \geq 3$, are independent parameters.
A more general form for $S_{irr}$
will be specified later, when we will choose the renormalization
conditions. Obviously a trivial choice is $S_{irr} = 0$.

Lattice perturbation theory is done using the Feynman rules, in which the
propagator is
\bea
 D(p) = \frac{1}{\hat{p}^2 + m^2}
\ena
and the vertices are contained in $S_I$.
The dependence of $D(p)$ on the ultraviolet cut-off $\Lambda_0$
(1.3) is not explicitly indicated.

The integrands of the Feynman integrals are periodic functions
of the momenta; it is not possible to define the degree of divergence
of a graph by considering its behavior for very large momenta, so that
the usual power-counting theorems on the continuum do not apply on the
lattice. A detailed analysis of the structure of the lattice diagrams
shows that it is nevertheless possible to define a lattice degree of
divergence and to establish a power-counting theorem on the lattice
\cite{r1}.

The fact that the $\phi^4$ model is perturbatively renormalizable
(multiplicatively)
on the lattice means that, for $n \geq 2$, the $n$-point connected
Green functions computed with these
Feynman rules admit finite limit, order by order in $\hbar$, in the
continuum limit $a \rightarrow 0$,~provided the constants $c_i^{(l)}$,
$i=1,2,3$ are chosen in such a way that the renormalization conditions
are satisfied, and provided the irrelevant
vertices contained in $S_{irr}$ are chosen to satisfy a bound to be
specified later. A proof of multiplicative renormalization of $\phi^4$
using BPHZ has been given by Reisz \cite{r1,r2}, where it was also
shown that certain irrelevant terms, like
$S_{irr}[\phi] = \sum_x \sum_{n=3}^{\infty} d_{2n} a^{2n} \phi_x^{2n}$,
do not change the renormalized Green functions.
Reisz has also shown that there is a considerable freedom in the choice
of the lattice propagator \cite{r1,r2,r3}. We will return to this
point at the end of this section.

We will give another proof of these facts
using an approach to renormalization started by Polchinski \cite{p1}
and improved in \cite{k2}.

The idea followed in \cite{k2} to study renormalization consists in
introducing a class of theories, labelled by a continuous parameter
$\Lambda \in [0, \Lambda_0]$, differing from the previous theory
only for the propagator, called $D_{\Lambda}$. For $\Lambda =0$
the two propagators are equal, $D_0(p) = D(p)$, while for
$\Lambda = \Lambda_0$ one has $D_{\Lambda_0} = 0$, so that for
$\Lambda = \Lambda_0$ perturbation theory is trivial, since the
proper vertices coincide with the bare vertices.
The Green functions of this class of theories satisfies a `heat' flow
equation, first considered by Wilson \cite{w3} and further studied
by Polchinski \cite{p1}.

Renormalization of the theory is studied investigating the `heat' flow
of the Green functions from $\Lambda = \Lambda_0$ to $\Lambda = 0$.
The theory at $\Lambda = 0$ is the one in which one is really
interested, that is the one with the action $(1.4-5)$.

Mixed boundary conditions are applied to determine this flow.
The renormalization conditions on the relevant vertices
(see e.g. the renormalization conditions $(1.7)$)
are interpreted as boundary conditions at $\Lambda = 0$; the
irrelevant vertices are fixed by the boundary conditions at
$\Lambda = \Lambda_0$.

Using the flow equations one can prove by induction inequalities
on the Green functions. From these inequalities one can show that the
Green functions are bounded, and finally that they converge, for
$\Lambda_0 \rightarrow \infty$.

In \cite{k2} one uses a smooth cut-off on the momenta, which vanishes below
an infrared scale $\Lambda$ and above an ultraviolet scale $\Lambda_0$.

On the lattice the situation is somewhat different, since the Feynman
integrands are periodic functions of the momentum, with periodicity
$2 \Lambda_0$ (see $(1.3)$ ). Some care must be taken in defining what
is a `small momentum' or a `large momentum'. The propagator of a field
is a periodic function on the Brillouin zone; roughly speaking, a momentum
is small if it is close to a maximum of the propagator; it is large
if it is close to a minimum of the propagator. The maximum of the
propagator $(1.8)$ for the scalar field is at $p_{\mu} = 0$ in the
Brillouin zone; due to periodicity it has infinite other maxima on
${\bf R}^4$, for
instance in $p_{\mu} = 2 \Lambda_0$.  On the other end the
point $p_{\mu} = \Lambda_0$ is a minimum of the scalar propagator.
Then $p_{\mu} = \Lambda_0$ is a `large momentum', while
$p_{\mu} = 2 \Lambda_0$ is a `small momentum'.
The notion of region of large momenta is substituted on the lattice
by the notion
of periodic band at a distance of the order of $\Lambda_0$
from the maxima of the propagators.

In the case of the scalar propagator, such a band can be identified
with a region around the boundary of the Brillouin zone.
In the case of staggered fermions, which we will discuss in a later section,
the propagator has a maximum in $p_{\mu} = 0$ and other
maxima along the border of the Brillouin zone,
so the band of large momenta does not coincide with the boundary of
the Brillouin zone.

We will show that there is a natural way to extend the flow equation
on the lattice; the infrared cut-off $\Lambda$ regulates the distance of
the periodic band mentioned above from the maxima of the propagators.
For $\Lambda = 0$ the band coincides with the
full Brillouin zone; as $\Lambda$ increases, the band recedes from the
maxima of the propagators of the fields, until for $\Lambda = \Lambda_0$
it becomes vanishingly small.

\vskip 0.5 cm
{\bf 1.2 The flow equation.}

\vskip 0.5 cm

Introduce an `infrared' cut-off $\Lambda$  in the propagator, with
$0 \leq \Lambda \leq \Lambda_0$,
\bea
D_{\Lambda}(p) = \frac{K_\Lambda(p)}{\hat{p}^2 + m^2}
\ena
where, for $\Lambda > 0$, $K_{\Lambda}(p)$ is a
$C^{\infty}$ function of
$\frac{\hat{p}^2}{\Lambda^2}$  which vanishes for
$\frac{\hat{p}^2}{\Lambda^2} \leq \frac{16}{\pi^2}$, is monotonously
increasing for $\frac{16}{\pi^2} < \frac{\hat{p}^2}{\Lambda^2} <
\frac{64}{\pi^2}$ and is equal to $1$ for
$\frac{\hat{p}^2}{\Lambda^2} \geq \frac{64}{\pi^2}$.
For $\Lambda = 0$ we define $K_0 \equiv 1$.

$K_{\Lambda}(p)$ is a periodic function on the Brillouin zone; for
$\Lambda$ small, it cuts off a slightly squeezed ball around $p = 0$.
For $\Lambda$ close to $\Lambda_0$, it is non-vanishing only in a
slightly squeezed ball around $p = (\frac{\pi}{a},...,\frac{\pi}{a})$.

This propagator satisfies the following conditions
\bea
D_{\Lambda_0} \equiv 0 ~~~~; ~~~~D_0(p) = \frac{1}{\hat{p}^2 + m^2}.
\ena
$D_{\Lambda}$ and $K_{\Lambda}$ depend on the
ultraviolet cut-off $\Lambda_0$; here and in the following the dependence
on $\Lambda_0$ will not be explicitly indicated.

Observe that
$lim_{\Lambda_0 \rightarrow \infty} D_{\Lambda}(p)$
is rotation ( $O(4)$ ) invariant.

The partition function is given by
\bea
Z_{\Lambda}[J] = N_{\Lambda} exp -\frac{1}{\hbar}
S_I(\hbar \frac{\delta}{\delta J})~
exp \frac{1}{2 \hbar}\int_p J(-p) D_{\Lambda}(p) J(p)
\ena
where the source $J(p)$ has the same support as $K_{\Lambda}(p)$.
$N_{\Lambda}$ is the normalization constant for which
$Z_{\Lambda}[0] = 1$.

Actually in the case of an infinite
lattice $N_{\Lambda}$ diverges; to avoid this infrared divergence
we could work on
a finite lattice as long as we deal with the generating functionals;
soon we will work with the $n$-point connected Green functions,
with $n \geq 1$, in which $N_{\Lambda}$, related to the vacuum diagrams,
does not appear. At that point we could take the limit
of infinite lattice. For notational semplicity we will work always with
an infinite lattice.

Expanding formally $S_I(\hbar \frac{\delta}{\delta J})$ in powers of
$\frac{\delta}{\delta J}$, and using the identity
\bea
e^{- \frac{1}{2 \hbar} \int_p J(-p) D_{\Lambda}(p) J(p)}
[ \prod_{i=1}^n \hbar \frac{\delta}{\delta J(-p_i)}]
e^{\frac{1}{2 \hbar} \int_p J(-p) D_{\Lambda}(p) J(p)} = ~~~~~~~~~~~~~~~~~
{}~~~~~~~~~~~~~~~~~~~~~\nonumber \\
\prod_{i=1}^n [\phi (p_i) + \hbar
D_{\Lambda}(p_i) \frac{\delta}{\delta \phi (-p_i)}]|_{\phi (p) =
D_{\Lambda}(p) J(p)}
= e^{ \frac{\hbar}{2} \int_p \frac{\delta}{\delta \phi (-p)}
D_{\Lambda}(p) \frac{\delta}{\delta \phi (p)} }
\prod_{i = 1}^n \phi (p_i) |_{\phi (p) = D_{\Lambda}(p) J(p)}
\nonumber
\ena
holding when $p_1,...,p_n$ are within the support of
$K_{\Lambda}(p)$, we get
\bea
Z_{\Lambda}[J] = N_{\Lambda}
e^{\frac{1}{2 \hbar} \int_p J(-p) D_{\Lambda}(p) J(p)}
e^{ \frac{\hbar}{2} \int_p \frac{\delta}{\delta \phi (-p)}
D_{\Lambda}(p) \frac{\delta}{\delta \phi (p)} }
e^{-\frac{1}{\hbar}S_I(\phi)}|_{\phi (p) = D_{\Lambda}(p) J(p)}
\ena
One has
\bea
Z_{\Lambda}[J] = e^{\frac{1}{\hbar}W_{\Lambda}[J]}
\ena
where $W_{\Lambda}[J]$ is the functional generator of the connected
Green functions. Substituting $(1.13)$ in $(1.12)$ one gets
\bea
e^{\frac{1}{\hbar}W_{\Lambda}[J] -
\frac{1}{2 \hbar} \int_p J(-p) D_{\Lambda}(p) J(p) }
= N_{\Lambda} e^{ \frac{\hbar}{2} \int_p \frac{\delta}{\delta \phi (-p)}
D_{\Lambda}(p) \frac{\delta}{\delta \phi (p)} }
e^{-\frac{1}{\hbar}S_I(\phi)}|_{\phi (p) = D_{\Lambda}(p) J(p)}
\ena
The RHS of this equation is the functional generator of the connected
Green functions, apart from the free contribution
$D_{\Lambda}(p)$ of the two-point function.
In the following we will study the amputated connected Green functions,
which are amputated of the propagators $D_{\Lambda}(p_i)$ attached
to the external legs. The functional generator for these Green
functions, apart from the $2$-point tree-level contribution,
will be called $V_{\Lambda}[\phi]$, and it satisfies
\bea
e^{+ \frac{1}{\hbar}V_{\Lambda}[\phi]} = N_{\Lambda}
e^{ \frac{\hbar}{2} \int_p \frac{\delta}{\delta \phi (-p)}
D_{\Lambda}(p) \frac{\delta}{\delta \phi (p)} }
e^{-\frac{1}{\hbar}S_I(\phi)}~.
\ena
$V_{\Lambda}[\phi]$ has the Volterra expansion
\bea
V_{\Lambda}[\phi] = \ln N_{\Lambda}  + \sum_{n=1}^{\infty} \frac{1}{n!}
\int_{k_1,...,k_n} \phi (-k_1)...\phi (-k_n)
(2\pi )^4 \delta_P^4(\sum_1^n k_i)
V_{\Lambda, n}(k_1,...,k_{n-1}) \nonumber
\ena
$V_{\Lambda, n}(k_1,...,k_{n-1})$
is symmetric under permutations of $k_1,...,k_n$, where
$k_n = -\sum_{i=1}^{n-1}k_i$, and it is periodic under
$k_{i \mu} \rightarrow k_{i \mu} + \frac{2\pi}{a}$
for any $i=1,...,n-1$ and for any $\mu$.

As in \cite{BM,b2} we will study the flow equation perturbatively in $\hbar$.
The amputated connected Green functions are defined as series in $\hbar$,
$V_{\Lambda, n} = \sum_{l \geq 0} \hbar^l V_{\Lambda, n}^{(l)}$.

Each graph contributing to $V_{\Lambda, n}^{(l)}(k_1,...,k_{n-1})$
is an amputated connected graph with internal propagators
$D_{\Lambda}$, and with the vertices contained in $S_I$.
While in $(1.14-15)$ we restricted the support of
$\phi (p)$ to be equal to the one of $K_{\Lambda}$, it is clear from
the Feynman graph representation that the external momenta can
be arbitrarily chosen.
$V_{\Lambda, n}^{(l)}(k_1,...,k_{n-1})$ is a $C^{\infty}$ function
on the Brillouin zone.

Differentiating equation $(1.15)$ with respect to $\Lambda$ one gets
the flux equation
\bea
\frac{\partial V_{\Lambda}[\phi]}{\partial \Lambda} =
\frac{1}{2} \int_p \frac{\partial D_{\Lambda}(p)}{\partial \Lambda}
[ \hbar \frac{\delta^2 V_{\Lambda}}{\delta \phi (-p) \delta \phi (p)}
+ \frac{\delta V_{\Lambda}}{\delta \phi (-p)}
\frac{\delta V_{\Lambda}}{\delta \phi (p)}] +
\hbar \frac{\partial \ln N_{\Lambda}}{\partial \Lambda}.
\ena

Differentiating $(1.16)$ with respect to $\phi (k_1),...,\phi (k_n)$
and setting $\phi$ to zero one gets the flux equations for the
amputated connected Green functions, for $n \geq 2$ and $l \geq 1$,
\bea
\frac{\partial V_{\Lambda, n}^{(l)}}{\partial \Lambda}(k_1,...,k_{n-1})
= \frac{1}{2} \int_p \frac{\partial D_{\Lambda}(p)}{\partial \Lambda}
V_{\Lambda, n+2}^{(l-1)}(k_1,...,k_{n-1},p,-p) ~~~~~~~~~~~~~~~\nonumber \\
+ \sum_P \sum_{n_1 + n_2 = n+2} \sum_{l_1 + l_2 = l}
\frac{\partial D_{\Lambda} }{\partial \Lambda}(\sum_{i=1}^{n_1-1} k_{Pi})
V_{\Lambda, n_1}^{(l_1)}(k_{P1},...,k_{P(n_1-1)} )
V_{\Lambda, n_2}^{(l_2)}(k_{Pn_1},...,k_{Pn} )
\ena
where $\sum_P$ is the sum over distinct permutations of
$k_1,...,k_n$, with $k_n = -\sum_1^{n-1} k_j$.

In the case $l = 0$ the first term in the RHS is absent, and for
$l= 0, n=2$ the zero solution must be chosen.

Since the sums in the RHS of $(1.17)$ satisfy $l_i \geq 0, ~n_i \geq 2$,
for given $l$ and $n$ they contain a finite number of terms and can
be solved iteratively. Indeed
in the RHS of $(1.17)$ one has $V_{\Lambda,E}^{(s)}$, with $s \leq l-1$,
or with $s=l$ and $E \leq n-2$. Ordering the vertices in such a way
that $V_{\Lambda,n_1}^{(l_1)}$ preceeds $V_{\Lambda,n_2}^{(l_2)}$
if $ l_1 < l_2$ or if $l_1=l_2$ and $n_1 < n_2$, it follows that in the RHS
of $(1.17)$ there appear only vertices preceeding $V_{\Lambda,n}^{(l)}$.
Therefore $(1.17)$ is a linear differential equation in
$V_{\Lambda,n}^{(l)}$.

$V_{\Lambda,n}^{(l)}$
can be determined assigning its value at a particular value of $\Lambda$.
At $\Lambda = \Lambda_0$ equation $(1.15)$ reduces to
\bea
V_{\Lambda_0}[\phi] = - S_I[\phi] + \hbar \ln N_{\Lambda_0}~~.
\ena
which provides a set of boundary conditions on $V_{\Lambda,n}^{(l)}$
at $\Lambda = \Lambda_0$. It follows that
at $\Lambda = 0$ one has
\bea
V_{0,2}^{(l)}(0) = \alpha^{(l)} ~~;~~
\frac{1}{8} \sum_{\mu = 1}^4 \frac{\partial^2 V_{0,2}^{(l)}}
{\partial p_{\mu} {\partial p_{\mu}}}(0) = \beta^{(l)} ~~;~~
V_{0,4}^{(l)}(0,0,0) = \gamma^{(l)}  \nonumber \\
\alpha^{(0)} = 0~~,~~ \beta^{(0)} = 0~~,~~ \gamma^{(0)} = -g~.~~~~~~~~~~~~~~~~
\ena

The flux equations for $n = 2$ and $n=4$ determine the three vectors
$(\alpha^{(l)}), (\beta^{(l)})$ and $(\gamma^{(l)})$
in terms of the three vectors $(c_i^{(l)})$ , $i=1,2$ and $3$
given in $(1.5)$;
in perturbation theory the viceversa is also true, since
in this case $(1.17)$
is a first-order linear differential equation.

Instead of fixing the boundary conditions for the flux equation $(1.17)$
on all the vertices at
$\Lambda = \Lambda_0$, one can equivalently impose the renormalization
conditions $(1.19)$ on the relevant terms, while keeping fixed
at $\Lambda = \Lambda_0$ the boundary conditions on the irrelevant
terms.

The renormalization conditions $(1.19)$ can be expressed in terms of the
renormalization conditions on the proper vertices $\Gamma_n$, using
the relations among the proper vertices and the connected Green functions.
A particular case is $\alpha^{(l)} = \beta^{(l)} = \gamma^{(l)} = 0$
for $l \geq 1$; in this case the renormalization conditions $(1.19)$
are equivalent to the renormalization conditions $(1.7)$.

{}From a practical point of view it is simpler to make perturbative
computations with proper vertices. However the flux
equations are simpler for the amputated connected  Green functions
than for the proper vertices, so we have chosen to prove
renormalizability using the former.
A proof of the renormalizability of $\phi^4$ using the flux equations
for the proper vertices has been given in \cite{b2}.

The Green functions
$lim_{\Lambda_0 \rightarrow \infty} V_{0, n}^{(l)}$
can be obtained from
$lim_{\Lambda_0 \rightarrow \infty} V_{\Lambda, n}^{(l)}$,
using $(1.15)$ and
the semi-group property of the heat kernel,
\bea
e^{+ \frac{1}{\hbar}V_{0}[\phi]} = \frac{N_0}{N_{\Lambda}}
e^{ \frac{\hbar}{2} \int_p \frac{\delta}{\delta \phi (-p)}
[ D_0 - D_{\Lambda} ](p) \frac{\delta}{\delta \phi (p)} }
e^{+ \frac{1}{\hbar}V_{\Lambda}[\phi]}
\ena
Following the same logic as in $(1.11-15)$, we can compute
$lim_{\Lambda_0 \rightarrow \infty} V_{0, n}^{(l)}$
in terms of Feynman graphs, in which the propagators are
given by
$lim_{\Lambda_0 \rightarrow \infty} [ D_0 - D_{\Lambda} ](p)$,
and the vertices are given by
$lim_{\Lambda_0 \rightarrow \infty} V_{\Lambda, n}^{(l)}$.

Since $m \neq 0$, the support of $[ D_0 - D_{\Lambda} ](p)$
is bounded; from this fact and from a bound on
$V_{\Lambda, n}^{(l)}$ which we will obtain later one could easily
show that,
if $lim_{\Lambda_0 \rightarrow \infty} V_{\Lambda, n}^{(l)}$
exists, then
$lim_{\Lambda_0 \rightarrow \infty} V_{0, n}^{(l)}$  exists.
Therefore we can fix the renormalization conditions at
$\Lambda = \Lambda_1$ instead of the conditions $(1.19)$:
\bea
V_{\Lambda_1,2}^{(l)}(0) = \alpha_1^{(l)} ~~;~~
\frac{1}{8} \sum_{\mu = 1}^4 \frac{\partial^2 V_{\Lambda_1,2}^{(l)}}
{\partial p_{\mu} {\partial p_{\mu}}}(0) = \beta_1^{(l)} ~~;~~
V_{\Lambda_1,4}^{(l)}(0,0,0) = \gamma_1^{(l)}
\ena
with $\alpha_1^{(0)} = \beta_1^{(0)} = 0$ and
$\gamma_1^{(0)} = - g$. The vectors
$\alpha^{(l)}, \beta^{(l)}$ and $\gamma^{(l)}$ can be computed in function
of the vectors $\alpha_1^{(l)}, \beta_1^{(l)}$ and $\gamma_1^{(l)}$;
order by order in perturbation
theory, the viceversa is also true. Therefore for $m \neq 0$ the
renormalization
conditions $(1.21)$ are equivalent to the renormalization conditions
$(1.19)$.  For technical reasons we will work with the former.
For $m = 0$ the renormalization conditions $(1.21)$
can be imposed, but it is not
trivial to prove the existence of the functional $V_{\Lambda}$ for
$\Lambda=0$, due to the presence of infrared divergences.
This issue is studied in \cite{b2,k4} using the flow equation.

\vskip 1.2 cm

{\bf 1.3 Proof of renormalizability.}

\vskip 0.5 cm

In the following we will use the symbol $P(x_1,...,x_n)$ to indicate
each time a different polynomial in $x_1,...,x_n$, with non-negative
coefficients independent of $\Lambda_0$.

Indicate formally by
$\partial_k^z V_{\Lambda, n}^{(l)}(k_1,...,k_{n-1})$ the differentiation of
$V_{\Lambda, n}^{(l)}(k_1,...,k_{n-1})$ $z$ times with respect to
$k_{i_1 \mu_1},..., k_{i_{n-1} \mu_{n-1}}$,
where $k_{i_j \mu_j}$ are external momenta.

Let $V(k_1,...,k_{n-1})$ be a differentiable function of the momenta
$k_1,...,k_{n-1}$.
In the following we will use the Taylor formula
\bea
V(k_1,...,k_{n-1}) = V(0,...,0) + \sum_{\mu}
\sum_{i=1}^{n-1} \int_0^1 dt k_{i \mu}
\frac{\partial}{\partial q_{i \mu }}
V(q_1,...,q_{n-1})|_{q_{i\mu}=tk_{i\mu}}
\ena

Consider a function $f_a(p_1,...,p_{n-1})$ which
is symmetric under permutations of $p_1,...,p_n$, where
$p_n = -\sum_{i=1}^{n-1}p_i$; the subscript $a$
indicates some collection of indices. Introduce the norm
\bea
|| f ||_{\Lambda} =
Max_{\{a\}} Sup_{M_{\Lambda}} |f_a(p_1,...,p_{n-1})|
\ena
where $M_{\Lambda}$ is the domain satisfying the conditions
$\hat{p}_i^2 \leq \frac{64}{\pi^2} Max(\Lambda^2, \eta^2)$,
for any $i=1,...,n$,
$p_n = - \sum_1^{n-1} p_i$;
$\eta$ is a fixed quantity, whose role will become clear later.
Since $|\hat{p}_{\mu}| \geq \frac{2}{\pi} | p_{\mu}|$
for $p_{\mu}$ belonging to the Brillouin zone, it follows that a
bound on $||f||_{\Lambda}$ implies a bound on the norm with
the form $(1.23)$, with the domain
$p_i^2 \leq 16 Max(\Lambda^2, \eta^2)$.
While the norm $(1.23)$ depends on $\Lambda_0$, the latter norm
does not.

An useful property of the norm $(1.23)$ is
$||f||_{\Lambda'} \leq ||f||_{\Lambda''}$ for
$\Lambda' \leq \Lambda'' \leq \Lambda_0$ .

Let us now discuss more precisely the conditions to be imposed
on the term
$S_{irr}$ in the action $(1.5)$.
We will assume that the coefficients $S_{\Lambda_0,n}^{(l)}(k_1,...,k_{n-1})$
of the perturbative Volterra expansion of $S_{irr}$
be $C^{\infty}$ functions on the Brillouin zone, satisfying the bounds
\bea
||\partial_k^z S_{\Lambda_0,n}^{(0)}||_{\Lambda} \leq
\frac{\Lambda^{5-n-z}}{\Lambda_0} P(\ln \frac{\Lambda_0}{\Lambda})~
,~l = 0 \nonumber \\
||\partial_k^z S_{\Lambda_0,n}^{(l)}||_{\Lambda} \leq
\frac{\Lambda^{5-n-z}}{\Lambda_0} P(\ln \frac{\Lambda_0}{\Lambda_1})~
{}~,~ l > 0
\ena
for any fixed $\eta$ and for
$\Lambda_1 \leq \Lambda \leq \Lambda_0$.
These conditions, together with the vanishing of
$V_{\Lambda_0,2}^{(0)} = S_{\Lambda_0,2}^{(0)}$,
give the previously mentioned conditions on $S_{irr}$.

Renormalization consists in showing that
$\partial_k^z V_{\Lambda,n}^{(l)}$ converges and is rotation-covariant
in the limit $\Lambda_0 \rightarrow \infty$. In order to prove
renormalizability we will first prove the following proposition.

For $\Lambda_1 \leq \Lambda \leq \Lambda_0$ one has
\bea
|| \partial_{\Lambda}^w \partial_y^v \partial_k^z
V_{\Lambda,n}^{(s)} ||_{\Lambda}
\leq \Lambda_0^{-v} \Lambda^{4-n-z-w+v}
P(\ln \frac{\Lambda}{\Lambda_1}, (\ln \frac{\Lambda_0}{\Lambda_1})^v )
\ena
for any $s$, for $v,w = 0, 1$,~ $z \geq 0$ and $n \geq 2$.
To prove this we will proceed by induction in the loop index.

Equation $(1.25)$ holds for $s = 0$, as can be seen by the following
power counting.

Consider a tree-level amputated connected graph with $n$ external legs,
$I$ internal propagators and $V_{\bar{n}}$ bare vertices with $\bar{n}$
legs; at tree level one
has $\bar{n} \geq 4$.

The internal propagators satisfy the bound
\bea
|| \partial_{\Lambda}^w \partial_k^z
 D_{\Lambda}(k) || \leq \alpha \Lambda^{-2-w-z}
\ena
where $\alpha$ is a positive constant.

This graph is bounded by
$const. \times \Lambda^b$, with
\bea
b = \sum_{\bar{n} \geq 4} (4-\bar{n}) V_{\bar{n}} - 2 I
\ena
where we used the bounds $(1.24)$,  $(1.26)$ and the fact that
$const. \times \Lambda^{4-n} > \frac{\Lambda^{5-n}}{\Lambda_0}
P(\ln \frac{\Lambda_0}{\Lambda})$.

$V = \sum_{\bar{n}} V_{\bar{n}}$
is the total number of vertices in the tree graph. Using the identities
\bea
n + 2 I = \sum_{\bar{n} \geq 4} \bar{n} V_{\bar{n}} ~~~;~~~ V - I =1
\ena
it follows that $b = 4 - n$.
Differentiating $V_{\Lambda,n}^{(0)}$ with respect to $y$, the only
non-vanishing contributions are those in which there is at least
one irrelevant
vertex $V_{\bar{n}}$, with $\bar{n} \geq 4$; using $(1.24)$, it follows
that
$|| \partial_y V_{\Lambda,n}^{(0)}||_{\Lambda}
\leq \frac{\Lambda^{5-n}}{\Lambda_0}
P(\ln \frac{\Lambda_0}{\Lambda})$ .

Using the fact that $K_{\Lambda}$ is a $C^{\infty}$ function on the
Brillouin zone,
it is easy to see that differentiating
the tree-level Green functions with respect to an external momentum,
or with respect to $\Lambda$, lowers $b$ by one.
Therefore the induction hypothesis is true for $s = 0$.

Convergence and rotation-covariance of
$\partial_k^z V_{\Lambda,n}^{(0)}$
for $\Lambda_0 \rightarrow \infty$ follow trivially.

Differentiating $(1.17)$ with respect to
$y$ and $k$ we get the following inequalities, for
$\Lambda_1 \leq \Lambda \leq \Lambda_0$,
\bea
|| \partial_{\Lambda} \partial_y^v \partial_k^z
V_{\Lambda,n}^{(l)} ||_{\Lambda} \leq d_1 ~\Lambda ~
|| \partial_y^v \partial_k^z V_{\Lambda,n+2}^{(l-1)} ||_{\Lambda} +
{}~~~~~~~~~~~~~~~~~~~~~~~~~~~~~~~~~~~~~~~~~~~~~~~~~~~~~~~~~\nonumber \\
\sum_{n_1+n_2 = n+2} \sum_{l_1 + l_2 = l}
\sum_{v_1+v_2=v} \sum_{z_1+z_2+z_3=z} d_{z_1,z_2,z_3} \Lambda^{-3-z_3}
|| \partial_y^{v_1} \partial_k^{z_1} V_{\Lambda,n_1}^{(l_1)} ||_{\Lambda}~
|| \partial_y^{v_2} \partial_k^{z_2} V_{\Lambda,n_2}^{(l_2)} ||_{\Lambda}~~~
{}~~~\ena
where $d_1$ and $d_{z_1,z_2,z_3}$
are constants, independent of $\Lambda$ and $\Lambda_0$.

Assume by induction hypothesis that $(1.25)$ holds for $s \leq l - 1$;
then we want to show that $(1.25)$ holds for $s=l$.

We will proceed in steps; in the first three steps we make a few
remarks which are used in the following steps.  Then we prove $(1.25)$
in a sequence of points, each point being established by using the previous
ones.  This sequence consists in increasing the number of legs and going,
for fixed number of legs, from irrelevant to relevant.  The first point
consists in proving $(1.25)$ for $s=l, n=2, z \geq 3$, that is for the
irrelevant part of the two-point function.

i) For $n + z \geq 5$ and $n > 2$ one has
\bea
|| \partial_y^v \partial_k^z V_{\Lambda,n}^{(s)} ||_{\Lambda} \leq
\int_{\Lambda}^{\Lambda_0} d \Lambda'
|| \partial_{\Lambda'} \partial_y^v \partial_k^z
V_{\Lambda',n}^{(s)} ||_{\Lambda'}
+ \frac{\Lambda^{5-n-z}}{\Lambda_0} P(\ln \frac{\Lambda_0}{\Lambda_1})
\ena
where we used the boundary conditions $(1.24)$ on the irrelevant operators.
Therefore if $(1.25)$
holds for $s=l, w=1$ and $n+z \geq 5$ , with $n > 2$, then $(1.25)$ holds for
$s=l$, $w=0$ and $n+z \geq 5$, $n > 2$.

ii) In the inequality $(1.29)$, the only terms in the RHS
in which there appear vertices of order $l$
are those with $n_1 \leq n-2$, $n_2 \geq 4$
or viceversa, since there are no tree-level two-point vertices.
If $(1.25)$ holds for $s=l$ up to
$n-2$ external legs, then by the induction hypothesis
it holds for $s=l$, with $n$ external legs and $w=1$.

iii) We will use the following inequality, holding for
$\Lambda_1 \leq \Lambda$,
\bea
|\partial_y^v \partial_k^z V_{\Lambda,n}^{(l)}(0,...,0)| \leq
|\partial_y^v \partial_k^z V_{\Lambda_1,n}^{(l)}(0,...,0)| +
\int_{\Lambda_1}^{\Lambda} d \Lambda' ||\partial_{\Lambda'}
\partial_y^v \partial_k^z V_{\Lambda',n}^{(l)}||_{\Lambda'}~~.
\nonumber
\ena

iv) The case of the two-point function can be treated in the
following steps:

a) using (ii) for $n=2$, and observing that there is no term
$V_{\Lambda,2}^{(l)}$ in the RHS of $(1.29)$,
it follows that $(1.25)$ holds for
$s=l$, $n=2$ and $w=1$;

b) from (iii) for $n=z=2$, $v=0$ and $\Lambda = \Lambda_0$,
and using the induction hypothesis,
it follows that the renormalization counterterms  satisfy the bound
$| c_2^{(l)} | \leq P(\ln \frac{\Lambda_0}{\Lambda_1})$.
Using this fact it follows that $(1.30)$ holds also for $s=l$ and
$n=2, z \geq 3$.
{}From (a) one gets $(1.25)$ for $s=l, n=2, z \geq 3$ and $w=0$;

c) making the Taylor expansion $(1.22)$ for
$\partial_y^v \partial_k^2 V_{\Lambda,2}^{(l)}(k)$ and using (iii)
it follows that, using the
renormalization conditions $(1.21)$, (a),(b), and the fact that in the
Brillouin zone one has
$k_{\mu} \leq \frac{\pi}{2} |\hat{k}_{\mu}|$,
we get $(1.25)$ for $s=l, n=z=2$ and $w=0$.

d) making the Taylor expansion $(1.22)$ for
$\partial_y^v \partial_k V_{\Lambda,2}^{(l)}(k)$,
using the hypercubic symmetry and (c) it follows that $(1.25)$ is true for
$s=l, n=2, z=1, w=0$.

e) making the Taylor expansion $(1.22)$ for
$\partial_y^v V_{\Lambda,2}^{(l)}(k)$,
using the renormalization conditions $(1.21)$, (a,d) and (iii),
it follows that $(1.25)$ is true for
$s=l, n=2, z=0, w=0$.

Therefore we have shown that the induction hypothesis
is true for $s=l, n=2$.

v) Use (i) for $n=4, z \geq 1$ to get $(1.25)$ for $s=l, n=4, z \geq 1, w=0$.
Using (iii) for $n=4, z \geq 0$ in the Taylor expansion of
$\partial_y^v V_{\Lambda,4}^{(l)}(k_1,k_2,k_3)$
we get, using the boundary conditions $(1.21)$ and the first part of (v),
the fact that $(1.25)$ holds
for $s=l, n=4, z=0, w=0$. Therefore the induction hypothesis
holds for $s=l, n=4$.

vi) Assume that $(1.25)$ is true for $s=l, ~4 \leq n' \leq n-2$; then using (i)
and (ii) it is easy to prove $(1.25)$ for $n$ external legs.
Since $(1.25)$ holds for $s=l, n=4$, then by induction it holds for
$s=l, n > 4$.

Therefore $(1.25)$ is true for $s=l$. By induction, it is true for any $s$.

\vskip 0.5 cm

{}From $(1.25)$ it follows that $V_{\Lambda, n}^{(l)}$ is bounded for
$\Lambda_0 \rightarrow \infty$; we have now to prove that
$V_{\Lambda, n}^{(l)}$ converges in this limit.

Define
$\tilde{D}_{z, \Lambda}(k) \equiv
lim_{\Lambda_0 \rightarrow \infty} \partial_k^z D_{\Lambda}(k)$
and
$\tilde{V}_{z,\Lambda, n}^{(l)}(k_1,...,k_{n-1}) \equiv
lim_{\Lambda_0 \rightarrow \infty}
\partial_k^z V_{\Lambda, n}^{(l)}(k_1,...,k_{n-1})$.

We will now prove the existence of
$\tilde{V}_{z,\Lambda, n}^{(l)}(k_1,...,k_{n-1})$;
moreover we will show that it is a continuous function of
the momenta and satisfies
\bea
\tilde{V}_{z,\Lambda, n}^{(l)}(k_1,...,k_{n-1}) =
\tilde{V}_{z, \Lambda_1, n}^{(l)}(k_1,...,k_{n-1})~~~~~~ ~~~~~~~~~~~~~~~~
\nonumber \\
+   \frac{1}{2} \int_{\Lambda_1}^{\Lambda}
d \Lambda' \{  \int \frac{d^4p}{(2\pi)^4}
\partial_{\Lambda'} \tilde{D}_{\Lambda'}(p)
\tilde{V}_{z,\Lambda', n+2}^{(l-1)}(k_1,...,k_{n-1},p,-p)
{}~~~~~~~~~~~~~~~~~~~~~~~~~~~~\nonumber \\
+ \hat{\sum}
\partial_{\Lambda'} \tilde{D}_{z_3, \Lambda'}(\sum_{i=1}^{n_1 -1} k_{Pi})
\tilde{V}_{z_1, \Lambda', n_1}^{(l_1)}(k_{P1},...,k_{P(n_1-1)} )
\tilde{V}_{z_2, \Lambda', n_2}^{(l_2)}(k_{Pn_1},...,k_{Pn} ) \} ~~~~~~~
\ena
for $n+z \leq 4$ and
\bea
\tilde{V}_{z, \Lambda, n}^{(l)}(k_1,...,k_{n-1}) =
 \frac{1}{2} \int_{\Lambda}^{\infty}
d \Lambda' \{  \int \frac{d^4p}{(2\pi)^4}
\partial_{\Lambda'} \tilde{D}_{\Lambda'}(p)
\tilde{V}_{z, \Lambda', n+2}^{(l-1)}(k_1,...,k_{n-1},p,-p)
\nonumber \\
+ \hat{\sum} \partial_{\Lambda'} \tilde{D}_{z_3, \Lambda'}
(\sum_{i=1}^{n_1 -1} k_{Pi})
\tilde{V}_{z_1 \Lambda', n_1}^{(l_1)}(k_{P1},...,k_{P(n_1-1)} )
\tilde{V}_{z_2, \Lambda', n_2}^{(l_2)}(k_{Pn_1},...,k_{Pn} ) \}~~~~~~~~~~
\ena
for $n+z \geq 5$,
Here and in the following of this section we will use the symbol
\noindent $\hat{\sum} =
\sum_P \sum_{n_1 + n_2 = n+2} \sum_{l_1 + l_2 = l}
\sum_{z_1 + z_2 + z_3 = z}$,
where $\sum_P$ is the sum over distinct permutations of
$k_1,...,k_n$, with $k_n = -\sum_1^{n-1} k_j$.

The proof of convergence is made by induction in the loop order $l$,
so that we start by considering the case $l=0$; in the limit
$\Lambda_0 \rightarrow \infty$ the only bare vertex
corresponds to the tree-level term  $g \phi^4$; the propagator
converges to $\tilde{D}_{\Lambda}$; a tree-level graph on the lattice
converges to the corresponding tree-level
graph of the continuum $g \phi^4$ theory and it is easy to check that
$(1.31-32)$ hold.

Assume that the induction hypothesis $(1.31-32)$ is true for $s < l$.
Then we want to show that it is true for $s=l$.
The integrated flux equations are
\bea
\partial_k^z V_{\Lambda, n}^{(l)}(k_1,...,k_{n-1}) =
\partial_k^z V_{\Lambda_1, n}^{(l)}(k_1,...,k_{n-1}) +~~~~~~~~~\nonumber \\
+   \frac{1}{2} \int_{\Lambda_1}^{\Lambda}
d \Lambda' \{ \int_p
\partial_{\Lambda'} D_{\Lambda'}(p)
\partial_k^z V_{\Lambda', n+2}^{(l-1)}(k_1,...,k_{n-1},p,-p)
{}~~~~~~~~~~~~~~~~~~~~~~~~~~~~\nonumber \\
+ \hat{\sum}
\partial_k^{z_3}\partial_{\Lambda'} D_{\Lambda'}(\sum_{i=1}^{n_1 -1} k_{Pi})
\partial_k^{z_1}V_{\Lambda', n_1}^{(l_1)}(k_{P1},...,k_{P(n_1-1)} )
\partial_k^{z_2}V_{\Lambda', n_2}^{(l_2)}(k_{Pn_1},...,k_{Pn} ) \}
\ena
for $n+z \leq 4$ and
\bea
\partial_k^z V_{\Lambda, n}^{(l)}(k_1,...,k_{n-1}) =
\partial_k^z V_{\Lambda_0, n}^{(l)}(k_1,...,k_{n-1})
{}~~~~~~~~~~~~~~~~~~~~~~~~~~~~~~~~~~~~\nonumber \\
+ \frac{1}{2} \int_{\Lambda}^{\Lambda_0}
d \Lambda' \{ \int_p
\partial_{\Lambda'} D_{\Lambda'}(p)
\partial_k^z V_{\Lambda', n+2}^{(l-1)}(k_1,...,k_{n-1},p,-p)
{}~~~~~~~~~~~~~~~~~~~~~~~~~~~~\nonumber \\
+ \hat{\sum}
\partial_k^{z_3} \partial_{\Lambda'} D_{\Lambda'}(\sum_{i=1}^{n_1-1} k_{Pi})
\partial_k^{z_1} V_{\Lambda', n_1}^{(l_1)}(k_{P1},...,k_{P(n_1-1)} )
\partial_k^{z_2} V_{\Lambda', n_2}^{(l_2)}(k_{Pn_1},...,k_{Pn} ) \}
\ena
for $n+z \geq 5$.

We start by proving convergence for $n=2, z \geq 3$. The RHS of $(1.34)$
depends only on lower perturbative order in the loop expansion,
so that by the induction hypothesis its vertices converge, and
the limit of the integrand exists. Furthermore, the first term in
the RHS of $(1.34)$ goes to zero in the limit.
{}From $(1.25)$ it follows that, for
$\eta \geq Max | k_i |$, the integrand appearing in the RHS of $(1.34)$
is bounded by a $\Lambda_0$-independent integrable function of the
integration variables; by the dominated convergence theorem of
Lebesgue, it follows that $(1.32)$ holds in this case.

For $n=2, z=2$ analogous considerations can be applied to the integral of the
RHS of $(1.33)$; the first term in the RHS of $(1.33)$ is
Taylor-expanded using $(1.22)$, using the
renormalization conditions $(1.21)$ and the convergence proof for
$n=2, z \geq 3$; it follows that
$\partial_k^2 V_{\Lambda, 2}^{(l)}$ converges.
The proof is repeated for $z=1$ and $z=0$ in a similar way.
For $n=4$ one realizes that the RHS of $(1.34)$ contains
$V_{\Lambda, n'}^{(l')}$  with $l'=l$ but only with $n' =2$ and so using
the previous result the proof holds again.

For $n > 4$, as in the proof of $(1.25)$, a systematic induction in
$n$ starts working with $(1.34)$.

Having proven that the vertices converge, let us now show that
they are continuous in the momenta. In fact, the integrands in the RHS of
$(1.31)$ and $(1.32)$ are continuous functions of the momenta by induction
hypothesis; using $(1.25)$ and $(1.26)$ for suitable $\eta$ we get
a momentum-independent integrable bound which is a function of the
integration variables, so that a standard theorem leads to the
conclusion that the RHS of equations $(1.31)$ and $(1.32)$ are continuous
functions of the momenta.

It is now an easy task to show that the limit for
$\Lambda_0 \rightarrow \infty$ and the limits involved in the
derivatives commute, that is
\bea
\tilde{V}_{z, \Lambda, n}^{(l)}(k_1,...,k_{n-1}) =
\partial_k^z lim_{\Lambda_0 \rightarrow \infty}
V_{\Lambda, n}^{(l)}(k_1,...,k_{n-1})
\ena
We give the proof for $z=1$.
{}From
\bea
V_{\Lambda, n}^{(l)}(k_1+\Delta k ,...,k_{n-1})
- V_{\Lambda, n}^{(l)}(k_1,...,k_{n-1})
= \int_{k_1}^{k_1+\Delta k} d p
\partial_p V_{\Lambda, n}^{(l)}(p,k_2,...,k_{n-1})
\ena
and by another application of the Lebesgue theorem we have
\bea
lim_{\Lambda_0 \rightarrow \infty}
\frac{V_{\Lambda, n}^{(l)}(k_1+\Delta k ,...,k_{n-1})
- V_{\Lambda, n}^{(l)}(k_1,...,k_{n-1})}{\Delta k} \nonumber \\
=\frac{1}{\Delta k}
\int_{k_1}^{k_1+\Delta k} d p
{}~lim_{\Lambda_0 \rightarrow \infty}
\partial_p V_{\Lambda, n}^{(l)}(p,...,k_{n-1})
\ena
where the integrand in the RHS is a continuous function, by the
previous results; taking the limit $\Delta k \rightarrow 0$
one gets immediately the assertion.

{}From $(1.25)$ it follows that
$lim_{\Lambda_0 \rightarrow \infty}
\partial_y V_{\Lambda, n}^{(l)} = 0$;
from the $y$-independence of the bound $(1.25)$ one could easily repeat
a proof analogous to $(1.35)$ of the fact that the two limits commute;
therefore the renormalized Green functions are independent of
$S_{irr}$. This independence is a manifestation of the large
arbitrariness in the discretization of the field theory on the lattice.

Rotation-covariance of
$lim_{\Lambda_0 \rightarrow \infty} V_{\Lambda, n}^{(l)}$
is proven inductively in the loop index, using the fact that
in $(1.31-32)$ the coefficient functions $\tilde{D}_{z, \Lambda}$
are rotation-covariant.

\vskip 1.0 cm

We have shown therefore that
$lim_{\Lambda_0 \rightarrow \infty} V_{\Lambda, n}^{(l)}(k_1,...,k_{n-1})$,
with the renormalization conditions $(1.21)$
are renormalized
rotation-covariant Green functions, satisfying the bound $(1.25)$,
order by order in the loop expansion; furthermore they are independent
of $S_{irr}$.
For $m \neq 0$ the Green functions
$lim_{\Lambda_0 \rightarrow \infty} V_{0, n}^{(l)}(k_1,...,k_{n-1})$,
with renormalization conditions $(1.19)$,
can be obtained from
$lim_{\Lambda_0 \rightarrow \infty} V_{\Lambda, n}^{(l)}(k_1,...,k_{n-1})$
using $(1.20)$ and therefore exist, are rotation-covariant and
are independent of $S_{irr}$.
This completes the proof of renormalizability.
\vskip 0.5 cm

{\bf 1.4 Generalizations.}
\vskip 0.5 cm

Let us now relax the condition required previously on
the tree-level two-point irrelevant terms.
It is easy to take into account irrelevant terms with two legs
at tree-level by including them in the kinetic term.
Add for example the term
\bea
\Delta S[\phi] = \frac{\xi}{2 \Lambda_0^2} \int_k
\phi (-k) (\hat{k}^2)^2 \phi (k)
\ena
to the action $(1.4-5)$. The propagator replacing $(1.8)$ is then
\bea
D_{\xi}(p) = \frac{1}{\hat{p}^2 + m^2 + \frac{\xi (\hat{p}^2)^2}{\Lambda_0^2}}
{}~. \ena
The parameter $\xi$ is chosen to satisfy
$\xi > - \frac{\pi^2}{16}$, in such a way that
\bea
|D_{\xi}(p)| \leq \frac{\beta}{\hat{p}^2 + m^2}
\ena
where $\beta$ is a positive constant.

In this case the propagator $D_{\xi}(p)$ has a single
maximum in the Brillouin zone, and it coincides with $D(p)$ in the
limit $\Lambda_0 \rightarrow \infty$.

The cut-off propagator is now defined as
$D_{\Lambda, \xi}(p) = K_{\Lambda}(p) D_{\xi}(p)$, replacing $(1.9)$.
Then we can repeat the proof of renormalization along the lines of
subsection $1.3$,
once we replace $D_{\Lambda}(p)$ with $D_{\Lambda, \xi}(p)$.
It should be noticed that the bound $(1.40)$ has been imposed in
order to verify the induction hypothesis $(1.25)$ at $s = 0$,
and that $\xi$ plays the role of $y$; for instance
from
\bea
||\partial_{\xi} D_{\Lambda, \xi} ||_{\Lambda} \leq
\frac{\alpha}{\Lambda_0^2}
\nonumber
\ena
it follows that $(1.25)$ holds for $s=0$, with $y$ replaced by $\xi$.
Therefore in the continuum limit the renormalized Green functions
do not depend on $\xi$.

A similar analysis holds for other two-point irrelevant terms.
One can prove renormalizability provided $(1.40)$ holds and the
propagator has the usual continuum limit \cite{r2,r3} .
\vskip 0.5 cm

A last remark on the irrelevant terms is in order. In the formulation
of subsection $1.3$,  $S_{irr}$ is an assigned function satisfying $(1.24)$.
In some application the irrelevant terms are instead assigned as a function
of the relevant terms, for instance one could be interested in an action
containing
\bea
c_3 \phi^4 + \frac{1}{\Lambda_0^2} c_3 \phi^6 + ...
\nonumber
\ena
To prove renormalizability the procedure is the one used in point (iv. b)
in subsection $1.3$.  One can easily show that the bound
\bea
| c_3^{(l)} | < P( \ln \frac{\Lambda_0}{\Lambda_1} )
\nonumber
\ena
holds. The rest of the induction
proof then follows as in the previous subsection.

\newpage

\sect{ Renormalization of matter field theories on a lattice.}

\vskip 0.5 cm

{\bf 2.1 Renormalization of matter field theories on a lattice.}

The renormalization of a general class of matter field
theories with a finite number of physical fields and of auxiliary fields
on a space-time lattice (not necessarily hypercubic) can be done in a similar
way
to the one followed in the first section. We will not reproduce all the
steps followed there; we will deal only with the points which are
significantly different in the two cases; we will restrict our attention
to rectangular lattices with lattice spacings
$a_1 \geq a_2 \geq a_3 \geq a_4$. We define
$\hat{p}_{\mu} = \frac{2}{a_{\mu}} sin \frac{p_{\mu} a_{\mu}}{2}$  and
$\Lambda_0^2 = \frac{\pi^2}{4} \sum_{\mu} a_{\mu}^{-2}$ .

Consider a lattice theory in which there are fields
$\Phi_i$ with canonical dimension $\Delta_i = 1, \frac{3}{2}$
or $2$, for scalars, fermions and auxiliary fields respectively.
For semplicity all fields are chosen to be real.
It is straightforward to generalize the analysis to the case of
complex fields.

The action reads, in momentum space,
\bea
S[\Phi] = \frac{1}{2} \int_p  \sum_{i,j} \Phi_i (-p) D^{-1}_{ij}(p)
\Phi_j (p) + S_I[\Phi]
\ena

The matrix propagator is a periodic $C^{\infty}$ function on the Brillouin
zone.

Introduce the propagator with infrared cut-off,
\bea
D_{\Lambda, ij }(p) = D_{ij}(p) K_{\Lambda}(p)
\ena

where $K_{\Lambda}$ is defined as after $(1.9)$.

 Assume that the propagator satisfies the following bounds:
\bea
||\partial_{\Lambda}^w \partial_k^z D_{\Lambda, ij} ||_{\Lambda}
\leq \alpha \Lambda^{\Delta_i + \Delta_j -4-z-w}
\ena
for $0 < \Lambda \leq \Lambda_0$, and $\alpha$ a positive constant
depending on $i,j,z$ and $w$.

$S_I$ contains the interaction terms, both relevant and irrelevant.
The tree-level part of $S_I$ is at least cubic in the fields;
each relevant tree-level term of $S_I$ has a parameter
which is independent from $\Lambda_0$.

The irrelevant terms are chosen to be linear in the parameter $y$.

As in the first section, introduce the functional  generator
$V_{\Lambda}[\Phi]$ for the amputated connected Green functions
(apart from the tree-level $2$-point contribution); each graph
contributing to these Green functions depends on $\Lambda$
only through its internal propagators $D_{\Lambda}$.
Therefore $V_{\Lambda}$ satisfies a flow equation
analogous to $(1.16)$.

Indicate with $V_{\Lambda,I}$, where $I = {i_1,...,i_n}$,
the amputated Green function with
external legs of fields $\Phi_{i_1},...,\Phi_{i_n}$.

The amputated connected Green functions are defined as series in $\hbar$,
$V_{\Lambda, I} = \sum_{l \geq 0} \hbar^l V_{\Lambda, I}^{(l)}$.

The flux equation has the form
\bea
\partial_{\Lambda} V_{\Lambda, I}^{(l)} = \frac{1}{2} \int_p \sum
\partial_{\Lambda} D_{\Lambda, jk} V_{\Lambda, I,jk}^{(l-1)}
+\sum \partial_{\Lambda} D_{\Lambda, jk}
V_{\Lambda, I_1,j}^{(l_1)} V_{\Lambda, I_2,k}^{(l_2)}
\ena
where $I_1 \cup I_2 = I$; for semplicity the momenta and the
sums have not been written explicitly.

Differentiating the flow equation with respect to $y$ and $k$
we get the following inequalities, for
$\Lambda_1 \leq \Lambda \leq \Lambda_0$,
\bea
|| \partial_{\Lambda} \partial_y^v \partial_k^z
V_{\Lambda,I}^{(l)} ||_{\Lambda} \leq \sum d_{ij}
\Lambda^{\Delta_i+\Delta_j-1}
|| \partial_y^v \partial_k^z V_{\Lambda,I,ij}^{(l-1)} ||_{\Lambda}
{}~~~~~~~~~~~~~~~~~~~~~~~~~~~~~~~\nonumber \\
+ \sum d_{z_1,z_2,z_3,I_1,I_2,i,j} \Lambda^{\Delta_i+\Delta_j-5-z_3}
|| \partial_y^{v_1} \partial_k^{z_1} V_{\Lambda,I_1,i}^{(l_1)} ||_{\Lambda}
|| \partial_y^{v_2} \partial_k^{z_2} V_{\Lambda,I_2,j}^{(l_2)} ||_{\Lambda}
{}~~~~~~~~~~~~\ena
where $d_{ij}$ and $d_{z_1,z_2,z_3,I_1,I_2,i,j}$
are constants, independent of  $\Lambda$ and $\Lambda_0$.

Impose the following renormalization conditions on the relevant terms:
\bea
\partial_k^z V_{\Lambda_1, I}^{(l)}(0,...,0) =
\alpha^{(l)}_{z, I}
\ena
for $\sum_{s=1}^n \Delta_{i_s} +z \leq 4$.

These renormalization conditions generalize those in $(1.21)$.
If there are symmetries in the theory, which can be maintained
at quantum level, one can impose relations among the renormalization
constants present in $(2.6)$; for instance in the case of the
scalar theory discussed in the first section, the hypercubic symmetry
and the discrete symmetry $\phi \rightarrow - \phi$ reduce the
conditions $(2.6)$ to the simpler form $(1.21)$.

Assume that $S_{irr}$ satisfies the following conditions:
\bea
||\partial_k^z S_{\Lambda_0,n}^{(l)}||_{\Lambda} \leq
\frac{\Lambda^{5-z-\sum_{s=1}^n \Delta_{i_s}}}{\Lambda_0}
P(\ln \frac{\Lambda_0}{\Lambda})~~,~~l = 0
\nonumber \\
||\partial_k^z S_{\Lambda_0,n}^{(l)}||_{\Lambda} \leq
\frac{\Lambda_0^{5-z-\sum_{s=1}^n \Delta_{i_s}}}{\Lambda_0}
P(\ln \frac{\Lambda_0}{\Lambda_1})~~,~~ l > 0 ~.
\ena
These equations generalize $(1.24)$.

\vskip 0.5 cm

Let us prove that for $\Lambda_1 \leq \Lambda \leq \Lambda_0$ one has
\bea
|| \partial_{\Lambda}^w \partial_y^v \partial_k^z
V_{\Lambda,I}^{(s)} ||_{\Lambda}
\leq \Lambda_0^{-v} \Lambda^{4- z-w+v-\sum_{r=1}^n \Delta_{i_r}}
P(\ln \frac{\Lambda}{\Lambda_1}, (\ln \frac{\Lambda_0}{\Lambda_1})^v)
\ena
for any $s$, for $v,w = 0, 1$,~ $z \geq 0$ and $n \geq 1$.

We will prove $(2.8)$ by induction in the loop index.

\vskip 0.5 cm

The assertion is true for $s = 0$, as can be seen by
the following power counting.

Consider a tree level amputated connected graph contributing to
$V_{\Lambda, I}^{(0)}$. Using the bounds $(2.3)$ and observing that
each factor $\Lambda^{\Delta_i}$ coming from an internal propagator
is cancelled by a factor coming from a vertex, it follows that
this graph is bound by
$const. \times \Lambda^b$, with
\bea
b = 4(V-I)- \sum_{ext} \Delta_i = 4 - \sum_{ext} \Delta_i
\ena
$\sum_{ext} \Delta_i$ is the sum of the canonical dimensions of
the fields appearing in the external legs of the graph.

Differentiating $V_{\Lambda,I}^{(0)}$ with respect to $y$, the only
non-vanishing contributions are those in which there is at least
one irrelevant vertex, satisfying $(2.7)$.
Therefore
$|| \partial_y^v V_{\Lambda,I}^{(0)}||_{\Lambda}
\leq \Lambda^{b+v} \Lambda_0^{-v} P(\ln \frac{\Lambda_0}{\Lambda})$.

It is easy to see that differentiating
the tree level Green functions with respect to an external momentum,
or with respect to $\Lambda$, lowers $b$ by one.
Convergence of
$\partial_k^z V_{\Lambda,I}^{(0)}$
for $\Lambda_0 \rightarrow \infty$ follows trivially.

Therefore the induction hypothesis is true for $s = 0$.

Assume that $(2.8)$ is true for $s \leq l - 1$;
then we will show that $(2.8)$ holds for $s=l$.

We can follow essentially the same steps as in the first section.
Let us only point out a few differences.

The first step is the same, provided $n+z \geq 5$ is substituted
by $\sum \Delta_i + z \geq 5$.

In the second step, in the inequality $(2.5)$ the only terms in the RHS with
$V_{\Lambda, I_1}^{(l)}$
are those with $n - 1$ external legs.
If $(2.8)$ holds for $s=l$ up to
$n - 1$ external legs, then it holds for $s=l$,
with $n$ external legs and $w=1$.

As in the first section, we prove $(2.8)$ by induction
in a sequence of points,
each point being established by using the previous ones.
This sequence consists in increasing the number of legs and going,
for fixed number of legs, from irrelevant to relevant.
It is easy to prove $(2.8)$ for the tadpoles.
The next point consists in proving $(2.8)$ for
$s=l, n=2$ and $\sum \Delta_i + z \geq 5$.
The rest of the proof is quite similar to the one in the scalar
case. Analogously one can prove convergence.

{}From $(2.8)$ it follows that
$lim_{\Lambda_0 \rightarrow \infty}
\partial_y V_{\Lambda, I}^{(l)} = 0$;
therefore the renormalized Green functions are independent of
$S_{irr}$.

If the relevant part of $S_I$ is hypercubic-invariant and the pointwise
limit of $D_{\Lambda ij}$ is $O(4)$-covariant in the limit
$\Lambda_0 \rightarrow \infty$, it is possible to choose the
renormalization conditions in such a way that the renormalized
Green functions are rotation covariant in the continuum limit.
The proof of this fact is along the lines of a similar proof in Section $1$.

\newpage

\sect{Renormalization of fermionic models.}
\vskip 0.5 cm

In this section we will study the renormalizability of Yukawa theories.
We will start by reviewing briefly the doubling problem and the
Wilson fermions. Since the Wilson propagator satisfies the bound
$(2.3)$, it is straightforward to prove the renormalizability of
Yukawa models with this kind of fermions, either using BPHZ
\cite{r1,r2,r3} or
the Polchinski approach here studied.

Staggered fermions \cite{s1} do not satisfy
the bound $(2.3)$, due to the presence of doublers at the border of the
Brillouin zone.
It has not yet been proven that interacting models
with staggered fermions are renormalizable. We describe a simple
kind of staggered fermions, which are staggered in one direction only,
and have a single doubler \cite{w4, p2}.
Although the fermionic propagator does not satisfy the bound $(2.3)$,
it satisfies a similar bound on a reduced Brillouin zone.
Using this fact, we study a chiral sigma model on a hypercubic lattice,
rewriting it equivalently
on a reduced Brillouin zone and showing that it is renormalizable
as a theory defined on a sublattice. Since renormalizability can be achieved
maintaining the same translation symmetries as in the theory defined on the
hypercubic lattice, renormalizability on the sublattice implies
renormalizability of the chiral sigma model
with this kind of staggered fermions on the hypercubic lattice.

The treatment of the flow equation with auxiliary fields, considered in the
previous section, plays a crucial role in this proof of renormalizability.

We conjecture that the renormalizability of Yukawa models with
Kogut-Susskind fermions can be proven in a similar way by studying
them first on a hypercubic sublattice with lattice spacing $2a$, on
which the fermionic propagator satisfies the bound $(2.3)$
\cite{k3} .

\vskip 0.5 cm

{\bf 3.1 Naive fermions, the doubling problem and Wilson fermions.}
\vskip 0.5 cm

The naive fermionic action for massless fermions on a space-time
hypercubic lattice is
\bea
I_0 = \frac{a^3}{2} \sum_x \sum_{\mu = 1}^4 {\bar{\psi}}_x \gamma_{\mu}
( \psi_{x + \hat{\mu}} -  \psi_{x - \hat{\mu}} )~;
\ena
the gamma matrices are hermitian and
satisfy $\{ \gamma_{\mu}, \gamma_{\nu} \} = 2 \delta_{\mu , \nu}$.

The inverse propagator for the naive fermionic action is
\bea
S^{-1}(p) = i \sum_{\mu=1}^4 \gamma_{\mu} \bar{p}_{\mu}
\ena
where $\bar{p}_{\mu} \equiv \frac{1}{a} sin(p_{\mu} a)$;
$S^{-1}(p)$ has
zeroes for $\sin p_{\mu}a = 0$, that is for $p_{\mu} = 0 , \frac{\pi}{a}$;
it describes $16$ Dirac fields in the continuum limit.
$I_0$ shares this property with any bilinear and
translationally invariant fermionic action,
whose propagator satisfies the following properties \cite{p3}:

i) reflection ($\Theta$ ) symmetry:
$S^{-1}(p_i,p_4) = \gamma_4 S^{-1\dag}(p_i,-p_4) \gamma_4$;

ii) hypercubic space-time lattice symmetry, i.e. invariance
under $\frac{\pi}{2}$ rotations of the coordinate axis;

iii) chiral symmetry:
$S^{-1}(p) = - \gamma_5 S^{-1}(p) \gamma_5$;

iv) locality, in the sense that $S^{-1}(p)$ is continuous
with its first derivatives.

{}From (i) and (ii) it follows that
$S^{-1}(p) = \gamma_{\mu} S^{-1\dag}(R_{\mu} p) \gamma_{\mu}$,
where $R_{\mu}$ is the reflection operator on the $\mu$-th
coordinate, $(R_{\mu}x)_{\nu} = (1 - 2 \delta_{\mu , \nu}) x_{\nu}$.
{}From this fact and (iii) it follows that $S^{-1}(p) = - S^{-1}(-p)$,
which together with periodicity $p_{\mu} \equiv p_{\mu} + \frac{2 \pi}{a}$
gives $S^{-1}(\bar{p}) = 0$
for $\bar{p}_{\mu} = 0 , \frac{\pi}{a}$; therefore a propagator satisfying
these conditions propagates $16$ modes.

This degeneracy is present also in the naive fermionic action for
massive fermions.

Wilson \cite{w2} eliminated this degeneracy introducing a term
which breaks the chiral symmetry in a hard way (since it has dimension $5$),
\bea
I_W = \frac{a^3 r}{2} \sum_x \sum_{\mu = 1}^4 \bar{\psi}_x ( 2 \psi_x
- \psi_{x + \hat{\mu}} - \psi_{x - \hat{\mu}} )~~.
\ena
The Wilson action is site-reflection positive for $r = 1$
\cite{l2} and it is link-reflection positive for
$ 0 < r \leq 1 $ \cite{o1} ( for a review see \cite{m1} ).
It describes one light mode and $15$ massive modes, with masses
of the order of $\frac{\pi}{a}$,
which decouple in the continuum limit.
\vskip 0.5 cm
The propagator for the Wilson fermion is, in the massive case,
\bea
S(p) =
\frac{-i \sum_{\mu} \gamma_{\mu}\bar{p}_{\mu} + M(p)}{
\bar{p}^2 + M^2(p)}
\ena
where
\bea
M(p) = M + \frac{r a}{2} \hat{p}^2
\ena
and $r$ is the Wilson parameter, satisfying $0 < r \leq 1$.
$M$ is the mass of the light fermionic mode.
We recall that
$\hat{p}_{\mu} = \frac{2}{a} sin \frac{p_{\mu}a}{2}$.

The fermionic propagator becomes the Dirac propagator in the
continuum limit $a \rightarrow 0$,
\bea
lim_{a \rightarrow 0} S(p) =
\frac{1}{i \sum_{\mu} \gamma_{\mu} p_{\mu} + M}
\ena
The denominator of $S(p)$ satisfies
the bound
\bea
\frac{1}{\bar{p}^2 + M^2(p)} \leq \frac{r^{-2}}{\hat{p}^2 + M^2}
\ena
where $r$ is the Wilson parameter. To prove this inequality,
observe that
\bea
\bar{p}^2 + M^2(p) = r^2 (\hat{p}^2 + M^2) +
(1-r^2) [ M^2 + \bar{p}^2 ]~~~~~~~~~~~~~~~~~~~~~~~~~~~~~~~
\nonumber \\
+ \frac{r^2a^2}{4} [ (\hat{p}^2)^2 - \sum_{\mu}\hat{p}_{\mu}^4 ]
+ raM \hat{p}^2 \geq r^2 ( \hat{p}^2 + M^2 )~~~~~~~~~~~~~~~~
\nonumber
\ena
for $0 < r \leq 1$ and $M \geq 0$.

It follows that the Wilson propagator satisfies the bound $(2.3)$.
Therefore, according to the previous section, one can construct
renormalizable models using Wilson fermions.
\vskip 1.0 cm

{\bf 3.2 ~ Fermions with minimal doubling.}

On a hypercubic  space-time lattice, translation-invariance, locality,
chiral symmetry and CP$\Theta$
(charge-conjugation $\times$ parity $\times$ reflection)
invariance of the action imply the existence of doubling
of the Dirac modes on the lattice \cite{k1}.

In order to have the minimal doubling allowed under these
assumptions, either reflection symmetry or hypercubic invariance
must be dropped.

In \cite{w4,p2} it has been given an example
of lattice translation invariant and chirally symmetric fermionic
action which breaks the hypercubic space-time symmetry to
cubic symmetry and which has minimal doubling.  The action is
\cite{p2}
\bea
I_f = I_0 + \frac{i a^3 \lambda}{2} \sum_x \sum_{\mu \neq 1} \bar{\psi}_x
\gamma_1 ( 2 \psi_x- \psi_{x + \hat{\mu}} - \psi_{x - \hat{\mu}})
+ M a^4 \sum_x \bar{\psi}_x \psi_x
\ena
where for $M \neq 0$ the chiral symmetry
\bea
\psi_x \rightarrow e^{i \beta \gamma_5} \psi_x~~~;~~~
\bar{\psi}_x \rightarrow \bar{\psi_x} e^{i \beta \gamma_5}
\ena
is softly broken, since $\bar{\psi}_x \psi_x$ has dimension $3$;
$I_0$ is the naive fermionic action (3.1).

The inverse propagator is
\bea
S^{-1}(p) = i \sum_{\mu} \bar{p}_{\mu} \gamma_{\mu}
+ \frac{i a}{2} \lambda \hat{p}_{\perp}^2 \gamma_1 + M.
\ena
where
$\hat{p}_{\perp}^2 \equiv \sum_{\mu \neq 1} \hat{p}_{\mu}^2$.
For $\lambda > 1/2$ there are only two propagating
 modes, around $p = (0,0,0,0)$ and $p = (\frac{\pi}{a},0,0,0)$ \cite{w4};
in fact the inverse propagator is equal to $M$ provided $\sin p_{\mu}a = 0$
for $\mu \neq 1$, which means $p_{\mu} = 0, \frac{\pi}{a}$ for
$\mu \neq 1$; and provided
$\sin p_1a + \lambda \sum_{\mu \neq 1} ( 1 - \cos p_{\mu}a ) = 0$
which cannot be satisfied if $\lambda > 1/2$ and $p_{\mu} = \frac{\pi}{a}$
for some $\mu \neq 1$. This is an example of fermions which are
staggered in one direction only.

For $\frac{1}{2} < \lambda \leq 1$ the action $(3.8)$ is also
link-reflection positive, so that one can construct a transfer matrix
over a Hilbert space of physical states \cite{p2} .

The hypercubic symmetry is broken to the cubic symmetry in the
directions $x_2, x_3$ and $x_4$, which includes the axis-inversion symmetry
$\psi_x \rightarrow i \gamma_{\mu} \gamma_5 \psi_{R_{\mu}x}$,
with $\mu \neq 1$.

The action $(3.8)$ has a discrete symmetry reflecting the fermion
in its mirror fermion:
\bea
\psi_x \rightarrow (-)^{x_1/a} \psi_{R_1x} ~~~;~~~
\bar{\psi}_x \rightarrow (-)^{x_1/a} \bar{\psi}_{R_1x}~~.
\ena
The action is link-reflection invariant, that is, invariant under
the antilinear mapping
\bea
\Theta \psi_{\underline{x},t} = \bar{\psi}_{\underline{x},1-t} \gamma_4
{}~~~~~;~~~ \Theta \bar{\psi}_{\underline{x},t} = \gamma_4
\psi_{\underline{x}, 1-t}
\ena
( it is also site-reflection invariant, but not site-reflection
positive ).
It is invariant under $CP$ transformations
\bea
\psi_{\underline{x},t} \rightarrow \gamma_4 C
\bar{\psi}_{\hat{1}-\underline{x},t}^T ~~;~~
\bar{\psi}_{\underline{x},t} \rightarrow
- \psi_{\hat{1}-\underline{x},t}^T C^{-1} \gamma_4
\ena
where $C$ is the charge conjugation matrix.
Therefore the action is CP$\Theta$-invariant. \\

The propagator $(3.10)$ does not satisfy the bound $(2.3)$,
since it is equal to $M^{-1}$ in the point $p = (\frac{\pi}{a},0,0,0)$.
Similar problems arise with the Kogut-Susskind staggered fermions
\cite{k3} .

This difficulty can be overcome redefining the fermionic variables
$\psi_x$ and $\bar{\psi}_x$, defined on the hypercubic lattice $\Xi$
in terms of new fermionic variables
defined on the sublattice $\Xi'$ characterized by $\frac{x_1}{a}$ even.
Define, for $\frac{x_1}{a}$ even,
\bea
\Psi_{1, x} = \frac{1}{2} (\psi_x + \psi_{x+\hat{1}}) ~~;~~
\bar{\Psi}_{1, x} &=& \frac{1}{2} (\bar{\psi}_x + \bar{\psi}_{x+\hat{1}})
\nonumber \\
\Psi_{2, x} = \frac{i}{2} \gamma_1 \gamma_5
(\psi_x - \psi_{x+\hat{1}}) ~~;~~
\bar{\Psi}_{2, x} &=&
\frac{i}{2} (\bar{\psi}_x - \bar{\psi}_{x+\hat{1}})
 \gamma_1 \gamma_5
\ena
This redefinition is similar to the one made in \cite{k3},
where the Kogut-Susskind
fermions are rewritten on a sublattice of lattice spacing $2a$.

The fermionic action $(3.8)$ becomes, in these variables,
\bea
I_F[\Psi, \bar{\Psi}] =
2 a^3 {\sum_x}'{\bar{\Psi}}_x [ \sum_{\mu \neq 1} \frac{1}{2}
\gamma_{\mu} \otimes 1
( \Psi_{x + \hat{\mu}} -  \Psi_{x - \hat{\mu}} )
+ \frac{1}{4} \gamma_1 \otimes 1
( \Psi_{x + 2\hat{1}} -  \Psi_{x - 2\hat{1}} ) ~~~~\nonumber \\
- \frac{i}{4} \gamma_5 \otimes \sigma_1
( 2 \Psi_x - \Psi_{x + 2\hat{1}} -  \Psi_{x - 2\hat{1}} )
+ i \frac{\lambda}{2} \sum_{\mu \neq 1}
\gamma_1 \otimes \sigma_3
( 2 \Psi_x - \Psi_{x + \hat{\mu}} -  \Psi_{x - \hat{\mu}} )
+ aM \Psi_x ] ~
\ena
where $\sum_x'$ is the sum over $x \in \Xi'$, and
where $\Psi_x = (\Psi_{i,x}), i=1,2$. $\sigma_i$ are the Pauli matrices.

The inverse propagator becomes
\bea
S^{-1}(p) = \Gamma + M =
{}~~~~~~~~~~~~~~~~~~~~~~~~~~~~~~~~~~~~~ \nonumber \\
i \{\sum_{\mu \neq 1} \gamma_{\mu} \otimes 1 \bar{p}_{\mu}
+ \frac{1}{2a} \gamma_1 \otimes 1 \sin (2 p_1a)
- a \gamma_5 \otimes \sigma_1 \bar{p}_1^2 +
\frac{\lambda a}{2} \gamma_1 \otimes \sigma_3
\hat{p}_{\perp}^2 \} + M
\ena
which is periodic on the reduced Brillouin zone
$|p_1| \leq \frac{\pi}{2a}$,
$|p_{\mu}| \leq \frac{\pi}{a},~ \mu \neq 1$.
In the continuum limit this propagator becomes the standard
Dirac propagator.

Let us show that this propagator satisfies the bound $(2.3)$.
Observe that
\bea
|| K_{\Lambda} S ||_{\Lambda} =
|| K_{\Lambda} \frac{\Gamma -M}{\Gamma^2 - M^2} ||_{\Lambda}
\leq c \Lambda ~
|| K_{\Lambda} \frac{1}{\alpha 1 + B} ||_{\Lambda}
\nonumber
\ena
where
\bea
\alpha = \bar{p}_1^2 + \hat{p}_{\perp}^2 + \frac{a^2}{4}
[ (\lambda \hat{p}_{\perp}^2)^2 - \sum_{\mu \neq 1} \hat{p}_{\mu}^4 ]
+ M^2
\nonumber
\ena
and
\bea
B = a \lambda  \hat{p}_{\perp}^2 ( 1 \otimes \sigma_3
\bar{p}_1 \cos p_1a - a i \gamma_1 \gamma_5 \otimes \sigma_2
\bar{p}_1^2 )~.
\nonumber
\ena
The smallest eigenvalue of $B$ is
$\beta = - a \lambda \hat{p}_{\perp}^2 |\bar{p}_1|$.

Therefore the minimum eigenvalue of $\alpha 1 + B$ is
\bea
\alpha + \beta \geq M^2 + \bar{p}_1^2 + \hat{p}_{\perp}^2
(1 - a \lambda |\bar{p}_1| ) + \frac{a^2}{4}
[ \lambda^2 (\hat{p}_{\perp}^2)^2 - \sum_{\mu \neq 1} \hat{p}_{\mu}^4 ] ~.
\ena

A detailed study of the RHS of $(3.17)$ shows that in the range
$\frac{1}{2} < \lambda \leq 1$
the following inequality holds:
\bea
\alpha + \beta \geq M^2 +
\frac{1}{5} (2 \lambda - 1)^2 (\bar{p}_1^2 + \hat{p}_{\perp}^2)~.
\nonumber
\ena

Using the norm $(1.23)$ and the inequalities
$\frac{4}{\pi^2} p^2 \leq \bar{p}_1^2 + \hat{p}_{\perp}^2 \leq p^2$,
for $p$ belonging to the reduced Brillouin zone, it follows that
\bea
|| K_{\Lambda} \frac{1}{\alpha 1 + B} ||_{\Lambda}
\leq const. \times \Lambda^{-2}
\ena
and therefore the propagator satisfies the bound $(2.3)$ for
$z = w = 0$ ; it is easy to see that $(2.3)$ is satisfied
for any $z$ and $w$.

The action $(3.15)$ has the link-reflection symmetry
\bea
\Psi_{\underline{x},t} \rightarrow
\bar{\Psi}_{\underline{x},1-t} \gamma_4 \otimes 1 ~~;~~
\bar{\Psi}_{\underline{x},t} \rightarrow
\gamma_4 \otimes 1 \Psi_{\underline{x},1-t}
\ena
It is invariant under the CP transformation
\bea
\Psi_{\underline{x},t} \rightarrow
\gamma_4 C \otimes \sigma_3 \bar{\Psi}_{- \underline{x},t} ~~;~~
\bar{\Psi}_{\underline{x},t} \rightarrow
\Psi_{- \underline{x},t}^T C^{-1} \gamma_4 \otimes \sigma_3
\ena
and it has the chiral symmetry (for $M = 0$ )
\bea
\Psi_x \rightarrow
e^{i\beta \gamma_5 \otimes \sigma_3} \Psi_x ~~~~;~~~~
\bar{\Psi}_x \rightarrow
\bar{\Psi}_x e^{i\beta \gamma_5 \otimes \sigma_3}
\ena
The discrete symmetry $(3.11)$ takes the form
\bea
\Psi_x \rightarrow \gamma_1 \gamma_5 \otimes \sigma_2 \Psi_{R_1x} ~~;~~
\bar{\Psi}_x \rightarrow
- \bar{\Psi}_{R_1x} \gamma_1 \gamma_5 \otimes \sigma_2
\ena
which mixes the $x_1$-inversion symmetry and the flavour symmetry.

\newpage

\vskip 0.5 cm

{\bf 3.3 Renormalizability of a chiral sigma model.}
\vskip 0.5 cm

Let us describe a sigma model on a hypercubic lattice,
with two charged fermions and two
neutral scalars, with a $U(1)_V$ vector symmetry and
a $U(1)_A$ axial symmetry.
The tree-level action is
\bea
I^{(0)} = I_f[\psi, \bar{\psi}] +
I_B[\sigma] + I_B[\pi] +
a^4 \sum_x [ g_1 \bar{\psi}_x ( \sigma_x + i \gamma_5 \pi_x ) \psi_x +
g_2 (\sigma_x^2 + \pi_x^2)^2 ]
\ena
where $I_f[\psi, \bar{\psi}]$ is the free fermionic action $(3.8)$
with $M=0$;

\bea
I_B[\phi] = \frac{a^2}{2} \sum_{x,{\mu}} \phi_x
( 2 \phi_x - \phi_{x + \hat{\mu}}
- \phi_{x - \hat{\mu}} ) +
a^4 \sum_x  \frac{m^2}{2}\phi^2_x
\nonumber
\ena
is the free bosonic action for $\phi = \sigma, \pi$~ .
$\sigma_x$ is a scalar, $\pi_x$ is a pseudoscalar.

The action $(3.23)$ has the $U(1)_V$ phase symmetry
$\psi_x \rightarrow e^{i\alpha} \psi_x$,~
$\bar{\psi}_x \rightarrow e^{-i\alpha} \bar{\psi}_x$;
the scalars are invariant under this symmetry. It has an axial
symmetry $U(1)_A$;
the chiral transformation on the fermions is given in $(3.9)$;
the scalars transform in the following way:
\bea
\sigma_x + i \pi_x \rightarrow e^{-2 i \beta}
(\sigma_x + i \pi_x)
\ena

The action $(3.23)$ has the discrete symmetry $(3.11)$ on the fermions, and
$\sigma_x \rightarrow \sigma_{R_1x}$,~
$\pi_x \rightarrow \pi_{R_1x}$ on the scalars;
it is invariant under CP and under cubic space-time rotations and
inversions in the $x_2 x_3 x_4$ directions.
It has the link-reflection symmetry defined by $(3.12)$
on the fermions, and by
$\Theta \sigma_{\underline{x},t} \rightarrow \sigma_{\underline{x},1-t}$,
{}~$\Theta \pi_{\underline{x},t} \rightarrow \pi_{\underline{x},1-t}$
on the scalars.

The fermionic propagator does not satisfy the bound $(2.3)$, so that
it is not straighforward to prove the renormalizability of this model.

As in the previous subsection, we will rewrite this model on an
anisotropic sublattice, in such a way that the fermion and its
doubler become two fermions which are coupled by an irrelevant term,
with propagator satisfying the bound $(2.3)$.
For consistency, it is also necessary to rewrite the scalars on this
sublattice; a scalar splits in a propagating mode and in an auxiliary
field.
Define, for $x_1$ even,
\bea
A_x = \frac{1}{2} (\sigma_x +\sigma_{x+\hat{1}})~~;~~
B_x = \frac{1}{2} (\pi_x +\pi_{x+\hat{1}}) \nonumber \\
F_x = \frac{1}{2} a^{-1} (\sigma_x -\sigma_{x+\hat{1}})~~;~~
G_x = \frac{1}{2} a^{-1}(\pi_x -\pi_{x+\hat{1}})
\ena
while the fermions are defined as in $(3.14)$.

The free $A-F$ part of the bosonic action $I_B[\sigma]$ becomes,
on the sublattice $\Xi'$,
\bea
I_{quadr}[A,F] = 2 a^4 {\sum_x}' \{ \frac{1}{4 a^2}
A_x (2 A_x - A_{x+2\hat{1}} -A_{x-2\hat{1}} )
+ \frac{1}{4} F_x (6 F_x + F_{x+2\hat{1}} +F_{x-2\hat{1}} )
\nonumber \\
+ \frac{1}{2a} F_x (A_{x+2\hat{1}} -A_{x-2\hat{1}} )
+ \frac{1}{2a^2} \sum_{\mu \neq 1} (
A_x (2 A_x - A_{x+\hat{\mu}} -A_{x-\hat{\mu}} ) ~~~~~~~~~~~~~~~~~\nonumber \\
+ a^2 F_x (2 F_x - F_{x+\hat{\mu}} -F_{x-\hat{\mu}} ))
+ \frac{m^2}{2}(A_x^2 + a^2 F_x^2) \} ~~~~~~~~~~~~~~~~~~~~~~
\ena
The interaction term is, at tree level,
\bea
I_{int}^{(0)} =2 a^4 {\sum_x}' \{ g_1 \bar{\Psi}_x
[ (A_x + i\gamma_1 \gamma_5 \otimes \sigma_1 a F_x )
+ i \gamma_5 \otimes \sigma_3
(B_x + i\gamma_1 \gamma_5 \otimes \sigma_1 a G_x )] \Psi_x \nonumber \\
+ g_2 [ ( A_x^2 + B_x^2 + a^2 F_x^2 + a^2 G_x^2 )^2
+4 a^2 (A_x F_x + B_x G_x )^2 ]  \}~~~~~~~~~~~~~~~~~~~~~
\ena
The tree-level action $(3.23)$ is equivalent to
\bea
I^{(0)} = I[A,F] + I[B,G] + I_F[\Psi, \bar{\Psi}] + I^{(0)}_{int}
\nonumber
\ena
The translation transformation by one site in the direction $x_1$
on $\Xi$ become, on $\Xi'$,
\bea
A_x \rightarrow \frac{1}{2} ( A_{x+2\hat{1}} + A_x ) + \frac{a}{2}
( F_{x+2\hat{1}} - F_x ) ~~~~~~~~~~~~~~~~~~~~~\nonumber \\
F_x \rightarrow -\frac{1}{2} ( F_{x+2\hat{1}} + F_x ) - \frac{1}{2a}
( A_{x+2\hat{1}} - A_x ) ~~~~~~~~~~~~~~~~\nonumber \\
\Psi_x \rightarrow 1 \otimes \sigma_3
[ \frac{1-i\gamma_1 \gamma_5 \otimes \sigma_1}{2} \Psi_x
+ \frac{1+i\gamma_1 \gamma_5 \otimes \sigma_1}{2} \Psi_{x+2 \hat{1}} ]
\nonumber \\
\bar{\Psi}_x \rightarrow [ \bar{\Psi}_x
\frac{1-i\gamma_1 \gamma_5 \otimes \sigma_1}{2}
+ \bar{\Psi}_{x+2 \hat{1}}
\frac{1+i\gamma_1 \gamma_5 \otimes \sigma_1}{2} ]
1 \otimes \sigma_3
\ena
and similarly for $B_x$ and $G_x$.
All these symmetries are preserved in presence of the cut-off
$K_{\Lambda}(p) = K(\frac{\hat{p}_{\perp}^2 + \bar{p}_1^2 }{\Lambda^2})$,
introduced for all the propagators.
Therefore they are compatible with the flow equation.
Choosing appropriately the renormalization conditions on the
relevant terms and the boundary conditions on the irrelevant terms,
it follows that the connected Green functions
$V_{\Lambda}^{(l)}$ have the translation symmetry $(3.28)$ for
any $\Lambda$.

Modulo irrelevant terms,
the most general quadratic scalar term in the sector $A-F$
of the bare action, which is invariant under the transformation $(3.28)$,
has the form
\bea
I_{A,F}^{quadr} =
2 a^4 {\sum_x}' \{ \frac{\alpha_1}{2}( A_x^2 + a^2 F_x^2) +
\frac{\alpha_2 + \alpha_3}{4 a^2}
A_x (2 A_x - A_{x+2\hat{1}} -A_{x-2\hat{1}} ) +
2 \alpha_2 F_x^2 \nonumber \\
+ \frac{\alpha_3 -\alpha_2}{4}  F_x (2 F_x - F_{x+2\hat{1}} -
F_{x-2\hat{1}} )
+ \frac{\alpha_2}{2a} F_x (A_{x+2\hat{1}} -A_{x-2\hat{1}} )
{}~~~~~~~~~~~~~~~~~\nonumber \\
+ \frac{\alpha_4}{2a^2} \sum_{\mu \neq 1} (
A_x (2 A_x - A_{x+\hat{\mu}} -A_{x-\hat{\mu}} )
+ a^2  F_x (2 F_x - F_{x+\hat{\mu}} -F_{x-\hat{\mu}} )) \}~~~~~~~~~~~~
\ena

Due to the axial symmetry the corresponding term in the $B-G$ sector
has the same coefficients $\alpha_1, \alpha_2, \alpha_3, \alpha_4$.

The same arguments lead to the quadratic fermionic term of
the bare action of the form
\bea
I_{\bar{\Psi} \Psi}^{quadr} =
2 a^4 {\sum_x}'{\bar{\Psi}}_x [ \beta_1 \sum_{\mu \neq 1} \frac{1}{2 a}
\gamma_{\mu} \otimes 1
( \Psi_{x + \hat{\mu}} -  \Psi_{x - \hat{\mu}} )
+ \beta_2 \frac{1}{4 a} \gamma_1 \otimes 1
( \Psi_{x + 2\hat{1}} -  \Psi_{x - 2\hat{1}} ) ~~~~\nonumber \\
- \frac{i}{4 a} \beta_2 \gamma_5 \otimes \sigma_1
( 2 \Psi_x - \Psi_{x + 2\hat{1}} -  \Psi_{x - 2\hat{1}} )
+ i \beta_3 \gamma_1 \otimes \sigma_3 \Psi_x + \beta_4 \Psi_x ] ~~~~~~~
{}~~~~~~~~~\ena

The bare interaction term has the same form as in $(3.27)$, with the
coupling constants $g_1$ and $g_2$ substituted by two bare parameters
$\chi_1$ and $\chi_2$.
Therefore starting with the bare lagrangian the renormalization
procedure will require $10$ independent conditions, at any loop order.

A parametrization of the two-point solution of the flux equation
in the $A-F$ sector
having the bare action just discussed as boundary condition is
\bea
V_{AA}^{\Lambda} = \Delta_{+}^{\Lambda}(p) +
 ~\Delta_{-}^{\Lambda}(p) \cos p_1a~~;~~
V_{FF}^{\Lambda} = a^2 (\Delta_{+}^{\Lambda}(p) -
\Delta_{-}^{\Lambda}(p) \cos p_1a ) \nonumber \\
V_{FA}^{\Lambda} = -i a  ~\Delta_{-}^{\Lambda}(p) \sin p_1a
\ena
where $\Delta_{+}^{\Lambda}(p)$ and $\Delta_{-}^{\Lambda}(p)$
have period $\frac{2\pi}{a}$ in the directions $p_2,p_3,p_4$
and are, respectively, periodic and antiperiodic in the direction
$p_1$, with period $\frac{\pi}{a}$.
They correspond to
$\frac{1}{4} [ V_{\sigma \sigma}^{\Lambda}(p_1,..) +
V_{\sigma \sigma}^{\Lambda}(p_1+\frac{\pi}{a},..) ]$ and
$\frac{1}{4} [ V_{\sigma \sigma}^{\Lambda}(p_1,..) -
V_{\sigma \sigma}^{\Lambda}(p_1+\frac{\pi}{a},..) ]$,
where $V_{\sigma \sigma}^{\Lambda}$ is the two-point
amputated connected Green function on the lattice $\Xi$.

Equation $(3.29)$ can be obtained from $(3.31)$ with a suitable choice
of $\Delta_{+}$ and $\Delta_{-}$ at $\Lambda = \Lambda_0$.

Equation $(3.31)$ leads to the following renormalization conditions
\bea
V_{AA}^{\Lambda_1 (l)}(0) = \alpha'^{(l)}_1~~;~~
\frac{d^2}{dp_1^2} V_{AA}^{\Lambda_1(l)}(0) = 4 (\alpha'^{(l)}_2
+ \alpha'^{(l)}_3 )~~;~~
\frac{d^2}{dp_k^2} V_{AA}^{\Lambda_1(l)}(0) = 2 \alpha'^{(l)}_4 \nonumber \\
V_{FA}^{\Lambda_1(l)}(0) = 0~~;~~
V_{FF}^{\Lambda_1(l)}(0) = 4 \alpha'^{(l)}_2 + a^2 \alpha'^{(l)}_1 ~~;~~
\frac{d}{dp_k} V_{FA}^{\Lambda_1(l)}(0) = 0~~, k \neq 1 \nonumber \\
\frac{d}{dp_1} V_{FA}^{\Lambda_1(l)}(0) = 2 i \alpha'^{(l)}_2
{}~~~~~~~~~~~~~~~~~~~~~~~~~~~~~~~~~~~~~~~~~~~~~~~~~~~~~~~~~~~~~\ena
Analogous renormalization conditions are imposed in the $B-G$ sector
 with the same coefficients $\alpha_1'^{(l)},...,\alpha_4'^{(l)}$.
The irrelevant terms in $(3.32)$ are kept to preserve the symmetry
$(3.28)$.

The renormalization conditions consistent with $(3.9)$, $(3.11-13)$,
$(3.28)$ and
with the axial reflection symmetry assign the
following expansion of $V_{\bar{\Psi} \Psi}$ in the neighborhood of $p =0$:
\bea
V_{\bar{\Psi} \Psi}^{\Lambda_1 (l)}(p) = 2 \beta_1'^{(l)} \sum_k p_k \gamma_k
+2 (\beta_2'^{(l)} p_1 + i \beta_3'^{(l)} ) \gamma_1 \otimes \sigma_3
+ O(p^2)
\ena
For the three-point functions the renormalization conditions are
\bea
V_{\bar{\Psi} \Psi A}^{\Lambda_1 (l)}(0) = \chi_1'^{(l)}~~;
V_{\bar{\Psi} \Psi B}^{\Lambda_1 (l)}(0) = i \gamma_5 \otimes
\sigma_3 \chi_1'^{(l)} ~~~~~~~~~~~~~~~~~~~~\nonumber \\
V_{A^4}^{\Lambda_1 (l)}(0) = \chi_2'^{(l)}~~;~~
V_{B^4}^{\Lambda_1 (l)}(0) = \chi_2'^{(l)}~~;~~
V_{A^2B^2}^{\Lambda_1 (l)}(0) = \frac{\chi_2'^{(l)}}{3}
\ena

As the fermionic propagators, also
the bosonic propagators associated to $(3.26)$ satisfy equation $(2.3)$,
with $\Delta_A = \Delta_B = 1$ and $\Delta_F = \Delta_G = 2$; therefore
the model is renormalizable on the lattice $\Xi'$.

\vskip 0.5 cm

{\bf 3.4 Back to the hypercubic lattice.}
\vskip 0.5 cm

We consider now the relation between the renormalization on the
cubic lattice $\Xi'$ and the hypercubic lattice $\Xi$.

This relation has some non-trivial feature already in the case of a
pure scalar theory. Let us consider a scalar theory on $\Xi$,
which is anisotropic in the $x_1$ direction. Then one must impose
three independent renormalization conditions on the two-point
function at every loop order. Considering instead the corresponding
theory on the lattice $\Xi'$, the scalar field is replaced by
a scalar field and an auxiliary field, as in subsection $3.3$;
one must impose four independent renormalization conditions,
as in $(3.32)$.
On $\Xi'$ there is one renormalization
condition which is not necessary on $\Xi$; in fact
$V_{FF}^{\Lambda_1 (l)}(0) = 4 \alpha'^{(l)}_2 + a^2 \alpha'^{(l)}_1$
corresponds to
\bea
V_{\Lambda_1, \phi \phi}^{(l)}(\frac{\pi}{a},0,0,0) =
\frac{8}{a^2} \alpha_2'^{(l)} + 2 \alpha_1'^{(l)}
\nonumber
\ena
This extra renormalization condition,  which is a requirement on the
behavior of the two-point function at the border of the Brillouin zone,
is allowed by the fact that
the $\alpha_2$-term and the $\alpha_3$-term
in the action $(3.29)$ are proportional, modulo an irrelevant term
which on $\Xi$ has the form
\bea
\int_p \phi (-p) \phi (p) (\hat{p}_1^2 - \bar{p}_1^2)~ .
\ena
Observe that this functional, which is irrelevant on $\Xi$,
is transformed into a relevant operator on $\Xi'$;
in fact the limit $p_1 \rightarrow \frac{\pi}{a}$ on $\Xi$ corresponds to
the limit $p_1 \rightarrow 0$ on $\Xi'$
and in this limit this operator does not vanish.
Therefore renormalizability of the scalar theory on $\Xi'$
is a stronger requirement than renormalizability on $\Xi$.

Let us show that for the chiral sigma model the renormalizability on
$\Xi'$ implies the renormalizability on $\Xi$.
Consider on $\Xi'$ a component of the Volterra expansion of $V$,
\bea
\int_{B/2}...\int_{B/2} \delta_{B/2}(\sum p)
\Phi_{I_1}...\Phi_{I_n} V_{I_1,...,I_n}
\ena
where $\Phi_I$ stands for the fields
$A,B,F,G,\Psi, \bar{\Psi}$. $\int_{B/2}$ stands for the integral on
Brillouin zone corresponding to $\Xi'$;
$\delta_{B/2}$ is the corresponding Dirac delta-function
of momentum conservation.
Using equations $(3.14)$ and $(3.25)$ in momentum space, the amputated
Green functions on $\Xi$ will be suitable linear combinations
of the amputated Green functions on $\Xi'$.
To show this, consider for example the following manipulation in $(3.36)$,
\bea
\int_0^{\frac{\pi}{a}} dp_1 ... \Psi_1 (p_1,...) =
2^{-\frac{1}{2}} \int_0^{\frac{\pi}{a}}..
[\psi (p_1,...)(1+e^{-ip_1a}) + \psi (p_1-\frac{\pi}{a},...)(1-e^{-ip_1a}) ]
\nonumber \\
= \int_{\frac{-\pi}{a}}^{\frac{\pi}{a}} dp_1 ...
\psi (p_1,...)(1+e^{-ip_1a})~~~~~~~~~~~~~~~~~~~~~~~~~~~~~~~~~
\ena
where we have used the periodicity of $\frac{\pi}{a}$ for the terms
not explicitly written. The integrals over the momenta $p_{\perp}$
is understood. Analogously
\bea
\int_0^{\frac{\pi}{a}} dp_1 ... \Psi_2 (p_1,...) =
2^{-\frac{1}{2}} \int_{-\frac{\pi}{a}}^{\frac{\pi}{a}} dp_1 ...
i \gamma_1 \gamma_5 \psi (p_1,...)(1-e^{-ip_1a})
\ena

\bea
\int_0^{\frac{\pi}{a}} dp_1 ... A(p_1,...) =
2^{-\frac{1}{2}} \int_{-\frac{\pi}{a}}^{\frac{\pi}{a}} dp_1 ...
\sigma (p_1,...)(1+e^{-ip_1a})
\ena

\bea
\int_0^{\frac{\pi}{a}} dp_1 ... F(p_1,...) =
2^{-\frac{1}{2}} \int_{-\frac{\pi}{a}}^{\frac{\pi}{a}} dp_1 ...
a^{-1} \sigma (p_1,...)(1-e^{-ip_1a})
\ena
and similarly for $B, G, \pi$.
Using $(3.37-40)$ for any $p_1$-integration variable, $(3.36)$
can be transformed in a multiple integral over the full Brillouin zone.

The delta-function of momentum conservation on the reduced Brillouin
zone can be expressed in terms of the delta-function on the full
Brillouin zone,
\bea
\delta_{B/2}(p_1,p_2,p_3,p_4) = \delta_B(p_1,p_2,p_3,p_4)
+ \delta_B(p_1+ \frac{\pi}{a} ,p_2,p_3,p_4)
\nonumber
\ena
This last formula shows that in general, due to the presence of the
translated delta-function, equations $(3.36-40)$ do not lead to the
Green functions of a theory on $\Xi$. But in our case, due to
the imposition of the symmetry $(3.28)$, this actually happens, indeed
in the sum all the terms containing the translated delta-function must
add to zero.
Therefore we can conclude that
every term $(3.36)$ leads to a  contribution to the Green
functions of the hypercubic lattice $\Xi$ convergent in the continuum
limit.

Moreover, $(3.38)$ shows that, concerning the fermionic lines,
not only the `naive limit' $a \rightarrow 0$ of the Green functions
on $\Xi$ at fixed momenta exists, but also the limit
$lim_{a \rightarrow 0} V_{\psi ..}(p_1 + \frac{\pi}{a},..)$ exists
and it is not vanishing; its
physical interpretation is of course the existence of a second
fermionic particle.

\vskip 0.5 cm

{\bf 3.5 $O(4)$ invariance.}
\vskip 0.5 cm

In order to get an easy proof of the $O(4)$ invariance of the theory
in the continuum limit, a further change of the field variables is suitable.
Indeed on $\Xi'$ the pointwise limit for $a \rightarrow 0$ of the
propagator in the scalar sector is not manifestly $O(4)$-invariant,
as can be seen
by Fourier-transforming equation $(3.26)$.

The amputated Green functions of the field $\sigma$
on $\Xi$ have contributions from both the Green functions with
$A$ and $F$ external legs; thus only in the sum, in the continuum limit,
the symmetry will be restored.

Let us make the redefinitions
\bea
A(x) = A'(x) ~~;~~
F(x) = F'(x) - \frac{1}{8a} (A'_{x+2\hat{1}} - A'_{x - 2\hat{1}} )
\ena
Analogously for the fields $B$ and $G$. In terms of the new fields
equation $(3.26)$ in momentum space becomes, up to irrelevant terms,
\bea
I [ A',F'] =
\frac{1}{2} \int_{B/2} \frac{d^4p}{(2\pi)^4} [ A'(p) A'(-p) (\hat{p}_{\perp}^2
+
\bar{p}_1^2 + m^2) + 4 F'(p)F'(-p) ]
\ena

Moreover $F'$ and $G'$ (as $F$ and $G$ before) appear in the interaction
part of the action only in the irrelevant terms.
{}From this fact one can easily show that in the continuum limit
the fields $F'$ and $G'$ decouple, in the sense that, up to the
$F'-F'$ and the $G'-G'$ two-point terms, all the interaction
amputated Green functions with some $F'$ and $G'$ external legs
vanish.
To show this, let us multiply by a parameter $y$ all the interaction
terms in the bare lagrangian which depend on $F'$ and $G'$.
For $y=1$ the action is equivalent to the one on the hypercubic
lattice $\Xi$, while for $y=0$ this is not true but the fields $F'$
and $G'$ trivially decouple.
In general, for $y \neq 1$, renormalization requires the imposition
of the renormalization conditions on all the terms with dimension lower
or equal to $4$. However we choose to impose again the conditions
corresponding to $(3.32), (3.33)$ and $(3.34)$.
The theory defined in this way is the original one for $y=1$;
after taking the limit $a \rightarrow 0$ the solution of the flow
equation does not depend on $y$, as it was shown in subsection $1.3$,
and in the remark in subsection $1.4$;
they are in particular equal to those of the $y=0$ case.

We showed in subsection $1.3$ that in the limit
$a \rightarrow 0$ the solutions of the flux equation satisfy the
system of integral equations $(1.31-32)$.
Now we have a system of equations involving functions with only
$A',B', \Psi, \bar{\Psi}$ external legs, while the two-point
functions of the auxiliary fields are  momentum-independent.

Only $O(4)$-invariant propagators
appear and then, as in subsection $1.3$, we conclude that with a
suitable choice of the constants of the renormalization conditions
(for instance all vanishing for $l \neq 0$ ) the continuum limit
of the solutions of the flux equations on $\Xi'$ are $O(4)$-covariant.
Thus from equations $(3.37-39)$, the terms on $\Xi$ coming from the
functions on $\Xi'$ with $A',B',\Psi, \bar{\Psi}$ legs will be
$O(4)$-covariant; from $(3.40)$ $V_{F'F'}$ could contribute
only to $V_{\sigma \sigma}$,
but an explicit computation shows that in the continuum limit this
contribution vanishes. Therefore $O(4)$-invariance on $\Xi$ is proven
for an arbitrary choice of the extra renormalization parameter
discussed in the previous subsection. The renormalized Green functions
do not depend on this parameter.

In this section we have studied a chiral sigma model, which has massless
particles; due to the infrared divergences
it is by no means trivial to show that
the existence of $V_{\Lambda_1}$ implies that of the functional $V$ at
$\Lambda = 0$. In the formalism of the flux equation (on the continuum)
this problem has been studied in \cite{b2,k4}.

\vskip 0.5 cm

{\bf  Conclusions.}
\vskip 0.5 cm

In this paper we have shown that the Polchinski approach to renormalization
can be applied on the lattice. Using this method we have proved the
renormalizability of the scalar and Yukawa models on the lattice
with Wilson fermions.
A proof of these facts has been obtained previously by Reisz \cite{r1,r2,r3}
using BPHZ.
We have also treated a Yukawa model with a simple kind of staggered fermions.
We proved that it is renormalizable provided a renormalization
condition on an irrelevant term is added to the usual renormalization
conditions;
this extra condition is related to the two-point scalar Green function at the
point on the border of the Brillouin zone in which there is the
doubler pole of the fermionic propagator.
While we have shown that such a renormalization procedure is sufficient to
prove the renormalizability, we have not shown that it is necessary.
Since the corresponding counterterm $(3.35)$ does not appear in the one-loop
bare action, only from a three-loop computation on the hypercubic lattice
one might see if this extra renormalization condition is necessary.

It would be interesting to
investigate the renormalizability of models
with Kogut-Susskind staggered fermions along the same lines on this paper.
For instance
in the case of four flavours (that is in the case of one naive fermion)
we expect that
a proof of renormalizability can be given,
involving fifteen extra renormalization conditions,
related to the two-point scalar Green function at the
points on the border of the Brillouin zone in which there are the
fermionic doublers.

\newpage


\vskip 0.5 cm

\begin{thebibliography}{99}

\bibitem{w1} K.G. Wilson, Phys. Rev. {\bf D10} (1974) 2445.

\bibitem{r1} T. Reisz, Comm. Math. Phys. {\bf 116} (1988) 81;
{\bf 116} (1988) 573.

\bibitem{r2} T. Reisz, Comm. Math. Phys. {\bf 117} (1988) 79;
{\bf 117} (1988) 639.

\bibitem{r3} T. Reisz, Nucl. Phys. {\bf B318} (1989) 417.

\bibitem{b1} N.N. Bogoliubov and D. Shirkov, Introduction to the theory
of quantized fields (Wiley/Interscience, New York).

\bibitem{r4} For a recent review see
V. Rivasseau, From perturbative to constructive renormalization
(Princeton University Press, Princeton, 1991).

\bibitem{n1} H.B. Nielsen and M. Ninomiya, Nucl. Phys. {\bf B185}
(1981) 20; Erratum {\bf B195} (1982) 541.

\bibitem{k1} L. H. Karsten, Phys. Lett. {\bf 104B} (1981) 315.


\bibitem{w2} K.G. Wilson, `Quarks and strings on a lattice', in
{\it New Phenomena in Subnuclear Physics}, ed. A. Zichichi, Part A (1975) 69.

\bibitem{l1} M. L\"{u}scher, `Selected topics in lattice field theory',
in {\it Fields, Strings and Critical Phenomena}, Les Houches (1988).

\bibitem{s1} J. Kogut and L. Susskind, Phys. Rev. {\bf D11} (1975) 395;\\
L.Susskind, Phys. Rev. {\bf D16} (1977) 3031.

\bibitem{w4} F. Wilczek, Phys. Rev. Lett. {\bf 59} (1987) 2397.

\bibitem{p2} M. Pernici, Phys. Lett. {\bf B346} (1995) 99.

\bibitem{p1} J. Polchinski, Nucl. Phys. {\bf B231} (1984) 269.

\bibitem{w3} K. G. Wilson, Phys. Rev. {\bf B4} (1971) 3174, 3184;\\
K. G. Wilson and J.G. Kogut, Phys. Rep. {\bf 12} (1974) 75.

\bibitem{k2} G. Keller, C. Kopper and M. Salmhofer, Helv. Phys. Acta
{\bf 65} (1992) 32; \\
R.D. Ball and R.S. Thorne, Ann. Phys. {\bf 236} (1994) 117.

\bibitem{BM} C. Becchi, `On the construction of renormalized quantum
field theory using renormalization group techniques',
in {\it Elementary particles, Quantum Fields and Statistical Mechanics},
eds. M. Bonini, G. Marchesini, E. Onofri, Parma University 1993.

\bibitem{b2}
M. Bonini, M. D'Attanasio and G. Marchesini, Nucl. Phys. {\bf B409}
(1993) 441.

\bibitem{k3} H. Kluberg-Stern, A. Morel, O. Napoly and B. Petersson,
Nucl. Phys. {\bf B220 [FS8]} (1983) 447.

\bibitem{p3} A. Pelissetto, Ann. Phys. (N.Y) {\bf 182} (1988) 177.

\bibitem{l2} M. L\"{u}scher, Comm. Math. Phys. {\bf 54} (1977) 283.

\bibitem{o1} K. Osterwalder and E. Seiler, Ann. Phys. (N.Y.) {\bf 110}
(1978) 440.

\bibitem{m1} I. Montvay and G. M\"{u}nster, Quantum fields on
a lattice (Cambridge University Press, 1994).

\bibitem{k4}  G. Keller and C. Kopper,
Comm. Math. Phys. {\bf 161} (1994) 515; \\
R.D. Ball and R.S. Thorne, `Infrared and ultraviolet behaviour of effective
scalar field theory' CERN-TH-7233-94.


\end{thebibliography}
\end{document}